\documentclass[twocolumn,twoside]{IEEEtran}

\ifCLASSINFOpdf

\else

\fi

\ifCLASSOPTIONcompsoc
\usepackage[caption=false, font=normalsize, labelfont=sf, textfont=sf]{subfig}
\else
\usepackage[caption=false, font=normalsize]{subfig}
\fi
\usepackage{lipsum}%
\usepackage[dvipsnames]{xcolor}

\usepackage{balance}
\usepackage{multicol}   
\usepackage{cite}
\usepackage{gensymb}
\usepackage{multirow}
\usepackage{graphics}
\usepackage{epsfig}
\usepackage{graphicx}
\usepackage{epstopdf}
\usepackage{textcomp}
\usepackage{amsmath}
\usepackage{mathtools}
\usepackage{filecontents}
\usepackage{lipsum,color}
\usepackage{amssymb}
\usepackage{float}

\usepackage{amsthm}
\newtheorem{remark}{Remark}

\usepackage{braket}

\usepackage{times} 
\usepackage{amsthm}  

\usepackage{amsfonts}

\newtheorem{proposition}{Proposition}

\theoremstyle{break}

\begin{document}
\title{Towards City-Scale Quantum Timing: Wireless Synchronization via Quantum Hubs}

\author{Mohammad~Taghi~Dabiri,~Mazen~Hasna, ~{\it Senior Member,~IEEE},
	~Rula~Ammuri, \\
	~Saif~Al-Kuwari,~{\it Senior Member,~IEEE},
	~and~Khalid~Qaraqe,~{\it Senior Member,~IEEE}
	\thanks{Mohammad Taghi Dabiri and Mazen Hasna are with the Department of Electrical Engineering, Qatar University, Doha, Qatar (e-mail: m.dabiri@qu.edu.qa; hasna@qu.edu.qa).}
	
	\thanks{Saif Al-Kuwari is with the Qatar Center for Quantum Computing, College of Science and Engineering, Hamad Bin Khalifa University, Doha, Qatar. email: (smalkuwari@hbku.edu.qa).}
	
	\thanks{Rula Ammuri is with Professionals for Smart Technology (PST), Amman, Jordan (email: rammuri@pst.jo).}
	
	\thanks{Khalid A. Qaraqe is with the College of Science and Engineering, Hamad Bin Khalifa University, Doha, Qatar, and also with the Department of Electrical Engineering, Texas A\&M University at Qatar, Doha, Qatar (e-mail: kqaraqe@hbku.edu.qa).}
	\thanks{This publication was made possible by NPRP14C-0909-210008 from the Qatar Research, Development and Innovation (QRDI) Fund (a member of Qatar Foundation), Texas A\&M University at Qatar, and Hamad Bin Khalifa University, which supported this publication. }
	
}

\maketitle
\begin{abstract}
This paper presents a novel wireless quantum synchronization framework tailored for city-scale deployment using entangled photon pairs and passive corner cube retroreflector (CCR) arrays. A centralized quantum hub emits entangled photons, directing one toward a target device and the other toward a local reference unit. The target, equipped with a planar CCR array, reflects the incoming photon without active circuitry, enabling secure round-trip quantum measurements for sub-nanosecond synchronization and localization.
We develop a comprehensive analytical model that captures key physical-layer phenomena, including Gaussian beam spread, spatial misalignment, atmospheric turbulence, and probabilistic photon generation. A closed-form expression is derived for the single-photon detection probability under Gamma-Gamma fading, and its distribution is used to model photon arrival events and synchronization error. Moreover, we analyze the impact of background photons, SPAD detector jitter, and quantum generation randomness on synchronization accuracy and outage probability.
Simulation results confirm the accuracy of the analytical models and reveal key trade-offs among beam waist, CCR array size, and background light. The proposed architecture offers a low-power, infrastructure-free solution for secure timing in next-generation smart cities.
\end{abstract}

%

%
\IEEEpeerreviewmaketitle


\section{Introduction}
Accurate timing and positioning are cornerstones of modern civilization. From coordinating financial transactions and managing power grids to enabling autonomous vehicles and seamless mobile communication, the synchronization of distributed devices across both space and time has become indispensable. Traditional solutions rely heavily on Global Navigation Satellite Systems (GNSS) and radio-based synchronization infrastructures, which are susceptible to spoofing, jamming, and environmental degradation, limiting their application to critical in dense urban environments or underground structures.

Emerging quantum technologies offer a fundamentally new paradigm for time and position coordination, enabling unprecedented levels of precision and security through quantum entanglement \cite{10487780}. In particular, the use of entangled photon pairs enables distributed nodes to perform highly accurate synchronization and ranging without relying on classical round-trip timing or electromagnetic wave propagation assumptions \cite{colombo2022time}.

\begin{figure}
	\begin{center}
		\includegraphics[width=3.35 in]{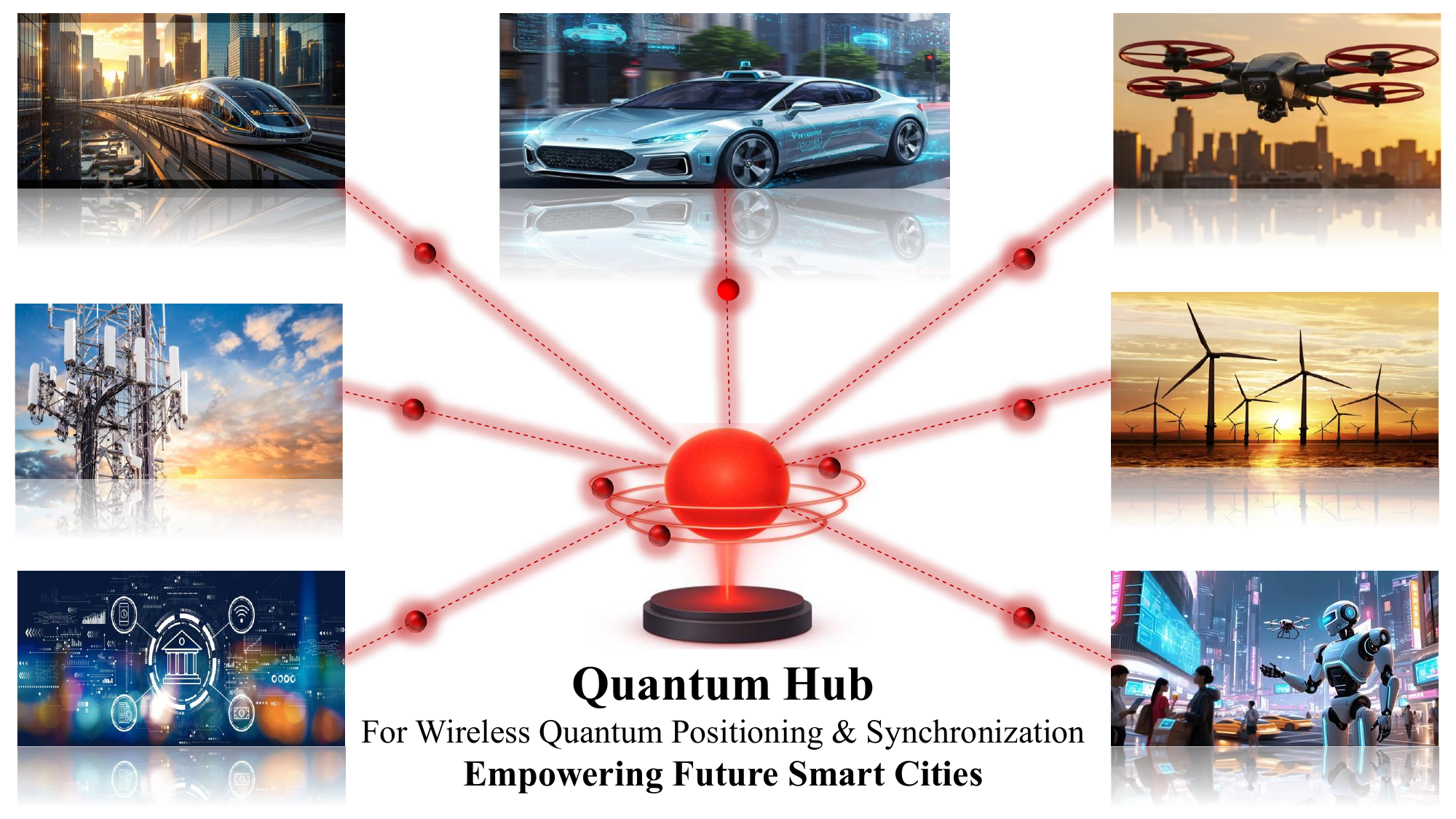}
		\caption{Conceptual illustration of a city-scale quantum synchronization infrastructure enabled by distributed quantum hubs. These hubs wirelessly synchronize and localize mobile and static devices (e.g., vehicles, UAVs, and sensors) across smart city environments using entanglement-assisted timing protocols. The architecture supports secure, infrastructure-free geolocation and sub-nanosecond synchronization across urban areas.}	
		\label{nv1}
	\end{center}
\end{figure}

\begin{figure*}
	\centering
	\subfloat[] {\includegraphics[width=3.35 in]{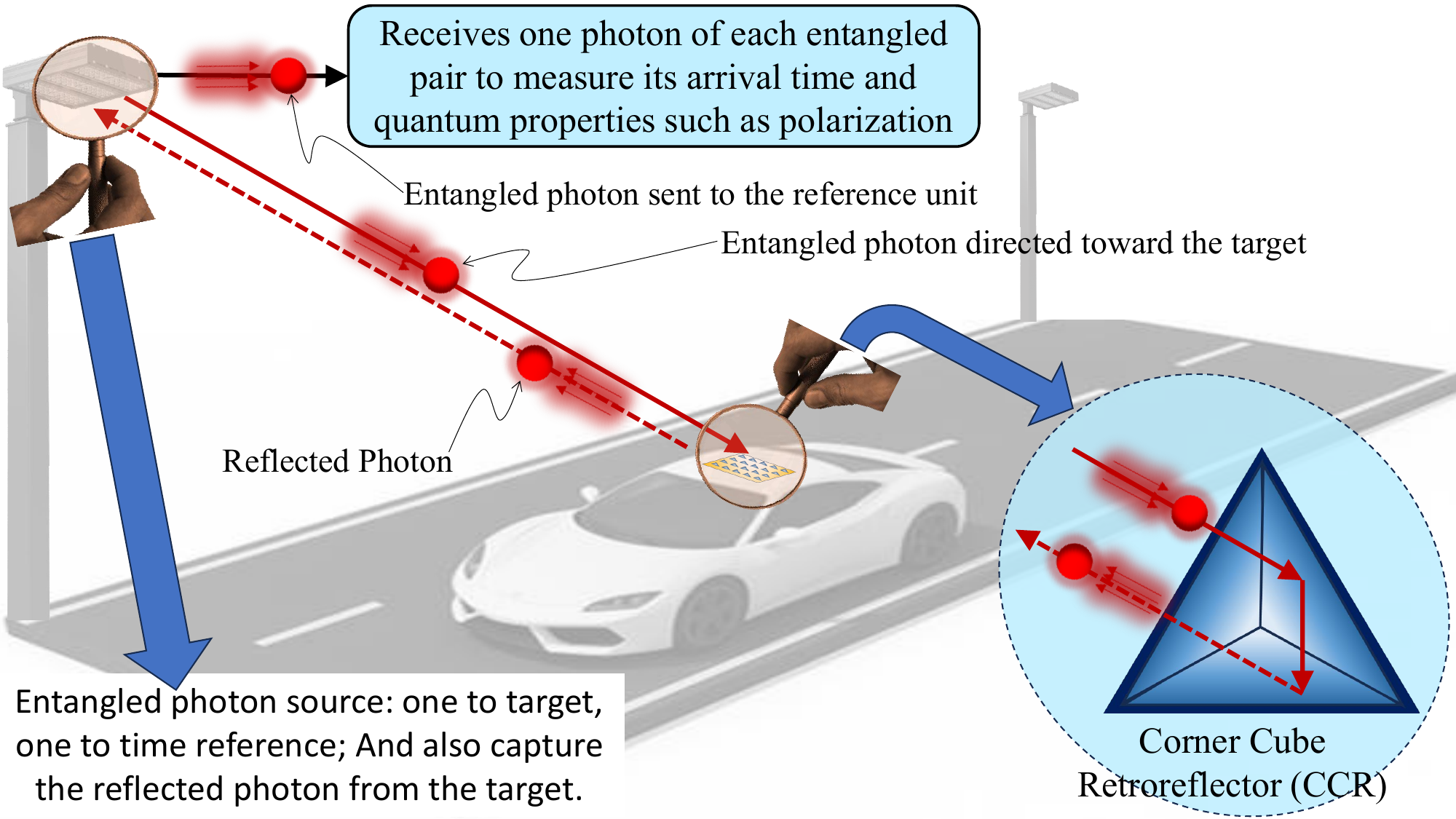}
		\label{cb1}
	}
	\hfill
	\subfloat[] {\includegraphics[width=3.35 in]{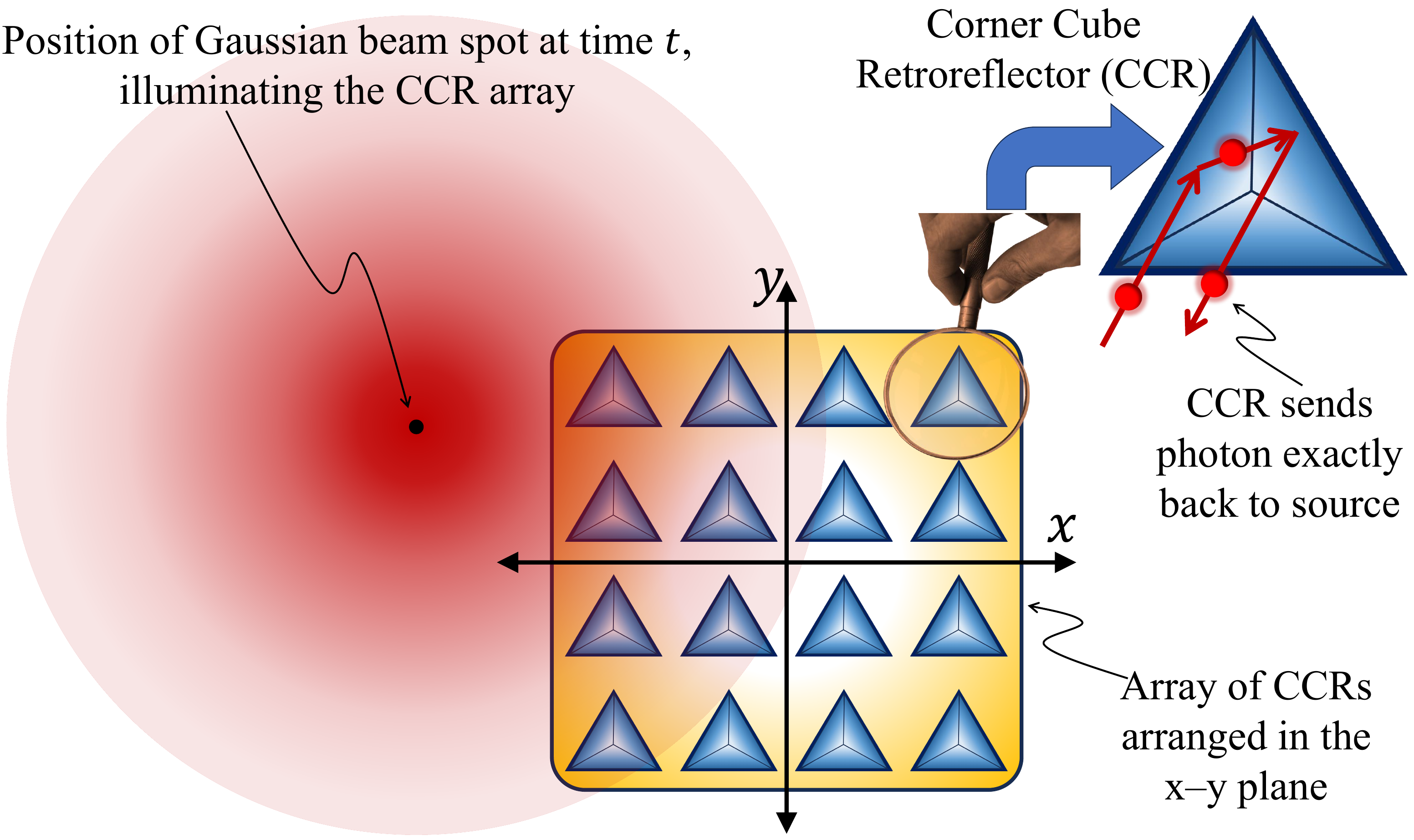}
		\label{cb2}
	}
	\caption{(a) System-level diagram of the proposed entanglement-based synchronization setup. A central quantum hub generates entangled photon pairs, directing one photon toward a passive CCR on the target and the other to a local time reference. The CCR reflects the incoming photon along the original path, allowing the hub to capture both the direct and reflected photons for coincidence-based timing analysis. (b) Geometry of the planar CCR array at the target side.}
	
	\label{cb}
\end{figure*}

In this work, we propose a city-scale wireless quantum synchronization architecture based on entangled photon pair distribution via centralized \textit{quantum hubs}. As illustrated in Figs.~\ref{nv1} and \ref{cb1}, a quantum hub transmits one photon of each entangled pair to a passive target device (e.g., a vehicle or sensor), while its twin photon is detected locally at the reference station. The target device is equipped with a planar corner cube retroreflector (CCR) array that reflects the photon exactly back along its incident path, enabling a round-trip quantum interaction without any active optical or electrical component on the target side. This not only drastically reduces power and hardware requirements at the endpoint but also allows for secure, high-resolution timing over long distances.

Such a framework opens avenues for a variety of futuristic applications, including:
\begin{itemize}
	\item Ultra-precise synchronization of vehicular fleets, aerial UAVs, and mobile robots in urban environments;
	\item Secure timestamping for financial, legal, or industrial blockchain systems;
	\item Infrastructure-free geolocation in GPS-denied areas such as tunnels, factories, or dense urban cores;
	\item City-scale quantum networks enabling entanglement-based sensor fusion, distributed clock synchronization, and secure communications.
\end{itemize}
We envision this new paradigm, quantum synchronization via wireless reflection and entanglement, to play a major role in next-generation smart cities, enabling scalable, low-power, and tamper-resistant coordination of billions of devices in a seamless and secure manner.

\begin{table*}[!t]
	\centering \textcolor{blue}{
	\caption{Comparison of Time Synchronization Approaches}
	\label{tab:comparison}
	\begin{tabular}{lccccc}
		\hline
		\textbf{Method} 
		& \textbf{Wireless} 
		& \textbf{Passive Target} 
		& \textbf{Urban Robustness} 
		& \textbf{Accuracy} 
		& \textbf{Remote HW} \\
		\hline
		GNSS-based \cite{zhu2022gnss,tavella2020precise,minetto2024nanosecond}
		& Yes & Yes & Low & $\sim$10--100 ns & Low \\		
		RF-based (PTP/NTP) \cite{patoli2025fpga}
		& Yes & No & Moderate & $\sim$100 ns--$\mu$s & Medium \\		
		Fiber-based (White Rabbit) \cite{gutierrez2021enhancing,sliwczynski2020picoseconds}
		& No & No & High & $<$1 ns & High \\		
		Quantum (Fiber / Hybrid) \cite{nande2025integrating,barhoumi2025quantum}
		& No & No & High & ps--sub-ns & High \\		
		Quantum (Free-space, active) \cite{spiess2023clock}
		& Yes & No & Limited & ps--sub-ns & High \\
		\hline
		\textbf{This work}
		& \textbf{Yes} & \textbf{Yes} & \textbf{High} & \textbf{sub-ns} & \textbf{Low} \\
		\hline
	\end{tabular}}
\end{table*}

Accurate time synchronization has long been a cornerstone of distributed systems, with solutions evolving from classical RF and satellite-based methods to advanced optical and quantum techniques \cite{chin2022digital,morales2019survey}. Each class of methods offers unique capabilities, but also presents critical limitations that motivate the need for our proposed architecture.

Global Navigation Satellite Systems (GNSS), such as GPS and Galileo, currently underpin most civilian and military time-distribution systems, offering synchronization accuracy on the order of 10–100~ns under open-sky conditions~\cite{zhu2022gnss,tavella2020precise,minetto2024nanosecond}. Complementary radio-based protocols, including the Network Time Protocol (NTP) and Precision Time Protocol (PTP), are widely adopted for local and wide-area network synchronization, delivering sub-microsecond accuracy in optimized Ethernet setups. Among them, IEEE~1588 (PTP) stands out due to its hardware timestamping capability and its adaptability to both wired and wireless environments. 
However, the precision of PTP is highly dependent on network topology, jitter, and delay asymmetry, which can potentially degrade synchronization to several hundred nanoseconds, especially over wireless or asymmetric links~\cite{patoli2025fpga}. Furthermore, both GNSS and radio-based methods remain vulnerable to intentional interference, such as jamming and spoofing, requiring continuous infrastructure support, which constrains their applicability in mobile, adversarial, or infrastructure-free scenarios.

To address the limitations of wireless and satellite-based systems, fiber-optic time transfer techniques have been developed, achieving synchronization precision to sub-nanosecond or even picosecond levels under controlled conditions~\cite{sliwczynski2020picoseconds}. By leveraging symmetric delay paths and hardware timestamping, protocols such as IEEE~1588 combined with Synchronous Ethernet, or White Rabbit extensions, have demonstrated timing accuracies below 1~ns over tens to hundreds of kilometers~\cite{gutierrez2021enhancing}. However, these systems require fixed physical infrastructure and careful calibration, which limits their deployment flexibility in dynamic or ad hoc environments.


Recent advances in entangled-photon generation and single-photon detectors have enabled proof-of-concept demonstrations of picosecond-class clock transfer over laboratory free-space links \cite{spiess2023clock}. 
Additionally, sub-nanosecond synchronization across three optical nodes using quantum nonlinear synchronization was experimentally demonstrated with thulium atoms in an optical cavity~\cite{nande2025integrating}. A hybrid quantum-classical framework using entangled spin qubits for sub-nanosecond synchronization and integration into IEEE~1588 has been proposed~\cite{barhoumi2025quantum}.

These quantum enabled approaches, while promising, remain largely limited to laboratory scale setups or specialized fiber infrastructures. Their applicability to real-world, dynamic environments, particularly those involving turbulent urban air and untethered mobile nodes, remains limited. Moreover, many proposed schemes rely on bulky, power-hungry, and alignment-sensitive hardware at both ends of the link, making them unsuitable for emerging applications that demand low-cost, compact, and maintenance-free endpoints, such as smart vehicles, wearables, or autonomous drones.
To date, no reported architecture tolerates strong urban-scale atmospheric fading, eliminates active optics or electronics at the remote node, and preserves the intrinsic security and sub-nanosecond precision of entanglement-based synchronization. 

\textcolor{blue}{Table~\ref{tab:comparison} summarizes and compares representative time-synchronization approaches from the literature, highlighting their key capabilities and limitations in terms of wireless operation, robustness, and deployment complexity.}

\subsection{Contributions}
Motivated by these limitations, this paper introduces a novel framework for wireless quantum synchronization tailored for city-scale deployments. By shifting all optical and computational complexity to a centralized hub, the proposed architecture enables fully passive, low-cost user nodes, making it well-suited for emerging applications such as mobile devices, autonomous platforms, and infrastructure-free quantum timing systems.
In particular, this paper introduces a novel framework for wireless quantum synchronization tailored for city-scale deployments. As generally shown in Fig. \ref{cb1}, the architecture combines a central \emph{quantum hub}, responsible for generating and transmitting entangled photons, with a \emph{planar CCR array} mounted passively on the remote target. By returning the partner photon along its original path without any onboard optics, electronics, or alignment control, the CCR inherently compensates for angular jitter and atmospheric fluctuations. To the best of our knowledge, this is the first wireless quantum synchronization scheme that simultaneously achieves turbulence robustness, eliminates remote-side infrastructure, and supports scalable, user-friendly operation in urban free-space environments. The key contributions are summarized as follows:
\begin{enumerate}
	\item \textbf{Quantum-enabled wireless architecture:} We model a new synchronization architecture in which a central quantum hub distributes entangled photons to passive mobile or static targets equipped with CCR arrays by taking into account the practical channel parameters. 
	
	\item \textbf{Spatially-resolved quantum channel modeling:} A detailed analytical model is developed to characterize the probability of photon detection in the presence of atmospheric turbulence, misalignment, aperture truncation, and retroreflection geometry. This includes a closed-form spatial probability expression and its statistical generalization.
	
	\item \textbf{Stochastic modeling of entangled photon measurements:} We formulate the entire entanglement-based detection process using structured bit and timestamp vectors derived from photon arrival events. Both signal and background photon processes are modeled under practical detector constraints, including SPAD jitter and time slot quantization.
	
	\item \textbf{Analytical synchronization error characterization:} A comprehensive analytical expression is derived for the synchronization timing error variance under realistic constraints. This model accounts for random signal photon counts, background contamination, and detector noise, and is validated via Monte Carlo simulations with excellent accuracy.
	
	\item \textbf{Outage probability and system design insights:} We provide tight analytical bounds on synchronization outage probability as a function of beam waist, CCR array density, and background photon rate. These results inform practical design trade-offs for urban quantum systems under both daylight and nighttime conditions.
\end{enumerate}

\begin{figure}
	\begin{center}
		\includegraphics[width=3.0 in]{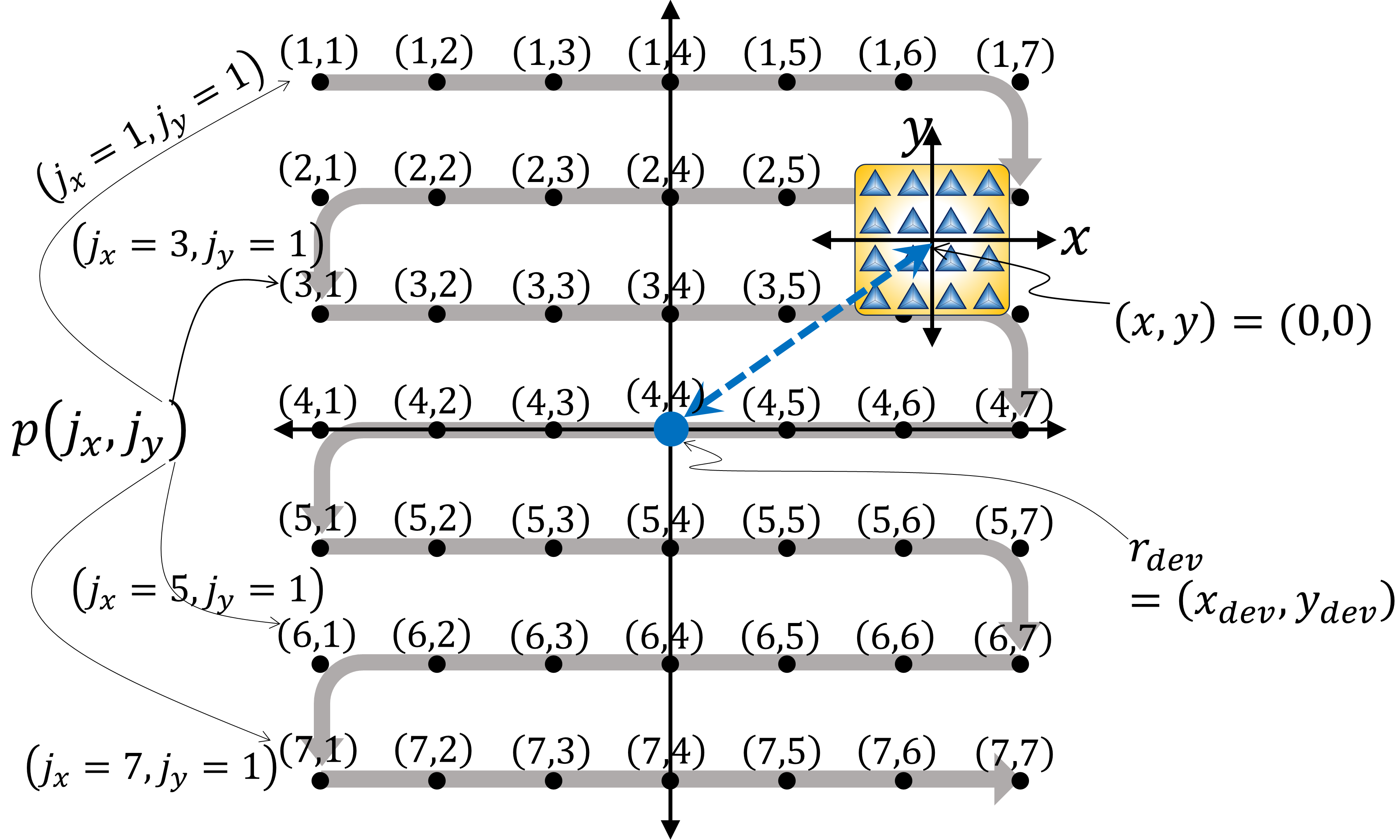}	
		\caption{Grid-based acquisition process around the initial position estimate $\mathbf{r}_{\mathrm{dev}} = (x_{\mathrm{dev}}, y_{\mathrm{dev}})$ of the CCR array center. The true target location is assumed to be at the origin $(0,0)$. The $x$--$y$ plane is divided into $N_{\mathrm{gr}} = N_{x}^{\mathrm{gr}} \times N_{y}^{\mathrm{gr}}$ cells, each centered at $\mathbf{p}(j_x, j_y) = \mathbf{r}_{\mathrm{dev}} + \mathbf{p}'(j_x, j_y)$. A Gaussian beam is sequentially directed to each cell for entangled photon probing and reflection detection.}
			
		\label{nb1}
	\end{center}
\end{figure}


\begin{table}[!t]
	\centering \textcolor{blue}{
	\caption{List of Key Notations}
	\label{tab:notations}
	\begin{tabular}{ll}
		\hline
		\textbf{Notation} & \textbf{Description} \\
		\hline
		$L_{\mathrm{tar}}$ & Target distance from the quantum hub \\
		$N_{\mathrm{ar}}$ & Total number of CCR elements \\
		$A_{\mathrm{ar}}$ & Effective area of a single CCR element \\
		$d_{\mathrm{ar}}$ & Spacing between adjacent CCR elements \\
		$r_{\mathrm{ap}}$ & Receiver aperture radius \\
		$\lambda$ & Optical wavelength \\
		$w_z$ & Beam waist at the target plane \\
		$\mathbf{r}_{\mathrm{dev}}$ & Spatial offset vector of the target \\
		$\sigma_p$ & Standard deviation of spatial misalignment \\
		$\sigma_\theta$ & Angular pointing/tracking error \\
		$C_n^2$ & Atmospheric turbulence strength parameter \\
		$\alpha,\beta$ & Gamma-Gamma turbulence parameters \\
		$\eta_{\mathrm{spad}}$ & SPAD detection efficiency \\
		$\sigma_{\mathrm{spad}}$ & SPAD timing jitter \\
		$t_{\mathrm{qb}}$ & Qubit time-slot duration \\
		$R_{\mathrm{qb}}$ & Photon-pair generation rate \\
		$\mu_t$ & Mean number of generated photon pairs per slot \\
		$\mu_{\mathrm{ch},j}$ & Mean received signal photon rate (grid $j$) \\
		$\mu_{\mathrm{bg}}$ & Mean background photon rate per slot \\
		$N_{\mathrm{sig}}$ & Number of detected signal photons \\
		$N_{\mathrm{bg}}$ & Number of detected background photons \\
		$N_{\mathrm{tot}}$ & Total detected photons ($N_{\mathrm{sig}}+N_{\mathrm{bg}}$) \\
		$t_{\mathrm{aq}}$ & Total acquisition duration \\
		$N_{\mathrm{gr}}$ & Number of spatial grid points \\
		$\mathbf{b}_{\mathrm{ref}}, \mathbf{b}_{\mathrm{rx}}$ & Reference and received bit vectors \\
		$\mathbf{t}_{\mathrm{ref}}, \mathbf{t}_{\mathrm{rx}}$ & Reference and received timestamp vectors \\
		$n_{\mathrm{ch}}$ & Synchronization estimation error \\
		$\mathbb{P}_{\mathrm{out}}$ & Synchronization outage probability \\
		$N_{\mathrm{t,min}}$ & Minimum required detections for synchronization \\
		\hline
	\end{tabular} }
\end{table}

\section{System Model}
We consider a quantum-enabled wireless synchronization architecture in which entangled photon pairs are generated and distributed for precise spatio-temporal tracking. One photon from each pair is sent to a reference station for timestamping and quantum-state characterization, while its entangled twin is transmitted to a target device that hosts a planar CCR array, as illustrated in Fig.~\ref{cb1}. This passive array reflects incident photons exactly back to their origin, allowing round-trip quantum measurements to be made wirelessly and without active components on the target side.

\subsection{CCR Array}
The target device to be quantum-synchronized and precisely localized is assumed to be positioned at an unknown radial distance $L_{\mathrm{tar}}$ away from the quantum source. Estimating $L_{\mathrm{tar}}$ is one of the main objectives of the proposed system. The target is equipped with a planar array of CCRs, as shown in Fig.~\ref{cb}, which ensures that any incident photon is reflected exactly back along its incoming path. This enables a precise two-way quantum measurement process without requiring active elements at the target.

To facilitate spatial modeling, we define a Cartesian coordinate system such that the center of the CCR array lies at the origin $(0,0)$ in the $x$--$y$ plane. The array is composed of $N_{\mathrm{ar}} = N_{\mathrm{ar}x} \times N_{\mathrm{ar}y}$ CCR units, where $N_{\mathrm{ar}x}$ and $N_{\mathrm{ar}y}$ denote the number of elements along the $x$ and $y$ axes, respectively. Each CCR occupies an effective area $A_{\mathrm{ar}}$ and is separated from adjacent units by a uniform pitch of $d_{\mathrm{ar}}$. Let $\mathbf{p}_{\mathrm{ar},i} = (x_{\mathrm{ar},i}, y_{\mathrm{ar},i})$ denote the $i$th CCR unit's spatial location in the $x$--$y$ plane, where $i \in \{1, \dots, N_{\mathrm{ar}}\}$.
This configuration results in a structured planar surface that can be scanned using a Gaussian beam for quantum measurements.

\subsection{Acquisition and Grid Scanning}
The transmitter initially relies on a coarse estimate of the target's location, represented by the spatial deviation vector $\mathbf{r}_{\mathrm{dev}} = (x_{\mathrm{dev}}, y_{\mathrm{dev}})$ in the $x$--$y$ plane where $(x_{\mathrm{dev}}, y_{\mathrm{dev}})\sim \mathcal{N}(0, \sigma_p^2) $ \cite{dabiri2022pointing, dabiri2018channel}. 
The first vector, denoted by $\mathbf{b}_\mathrm{ref}$, contains the logical bit value (0 or 1) inferred from the photon’s polarization state; for example, horizontal (H) or vertical (V). Accurate estimation of both $\mathbf{r}_{\mathrm{dev}}$ and the radial distance $L_{\mathrm{tar}}$ is critical for precise quantum localization.

To refine this initial estimate, the system performs a structured acquisition scan around $\mathbf{r}_{\mathrm{dev}}$. A rectangular search grid is defined in the $x$--$y$ plane centered on $\mathbf{r}_{\mathrm{dev}}$, partitioned into $N_{\mathrm{gr}} = N_{\mathrm{gr}x} \times N_{\mathrm{gr}y}$ discrete cells. Each cell in the grid is indexed by $(j_x, j_y)$ with $j_x \in \{1, \dots,  N_{\mathrm{gr}x}\}$ and $j_y \in \{1, \dots,  N_{\mathrm{gr}y}\}$, and its center is denoted by $\mathbf{p}(j_x, j_y)$. The cells of the grid are uniformly spaced, with adjacent centers separated by a fixed step size $d_{\mathrm{gr}}$.
We define $\mathbf{p}'(j_x, j_y)$ as the ideal relative location of the $(j_x, j_y)$th grid cell with respect to the grid center, given by:
\[
\mathbf{p}'(j_x, j_y) = \left( \left( j_x - \frac{ N_{\mathrm{gr}x} + 1}{2} \right) d_{\mathrm{gr}}, 
\left( j_y - \frac{ N_{\mathrm{gr}y} + 1}{2} \right) d_{\mathrm{gr}} \right),
\]
where $j = (j_y - 1) \cdot  N_{\mathrm{gr}y} + j_x$.
The absolute center location of each grid cell $\mathbf{p}(j_x, j_y)$ can then be expressed as:
\begin{align}
	\mathbf{p}(j_x, j_y) 
	&= \mathbf{r}_{\mathrm{dev}} + \mathbf{p}'(j_x, j_y). \label{eq:gridcell_position}
\end{align}

At each discrete time slot $t_j$, where $j \in \{1, \dots, N_{\mathrm{gr}}\}$, the entangled photon transmitter directs a narrow Gaussian beam toward the center of one grid cell. This process continues until all $N_{\mathrm{gr}}$ grid points have been probed, forming a complete scan cycle. The total acquisition time is given by
\begin{align}
	t_{\mathrm{aq}} = N_{\mathrm{gr}} \cdot t_j,
	\label{eq:taq}
\end{align}
where $t_j$ is the duration allocated for each beam steering operation.

\subsection{Quantum Transmitter Modeling}
In this work, we employ a two-level polarization encoding using horizontal and vertical (H/V) polarization states to represent logical bits 0 and 1. For FSO channels, this basis offers a practical trade-off between polarization stability and simplicity of implementation, enabling high-rate photon-pair generation and detection with minimal complexity \cite{savino2024robust}.
The quantum transmitter is responsible for generating time-correlated entangled photon pairs used in the positioning and synchronization process. These entangled pairs are produced at a rate of $R_{\mathrm{qb}}$ pairs per second, corresponding to a temporal resolution of $t_{\mathrm{qb}} = \frac{1}{R_{\mathrm{qb}}}$ for each quantum bit (qubit). Ideally, the generation of each photon pair occurs at the midpoint of the time slot, i.e., at $t_{\mathrm{qb}}/2$ within each interval.

In practice, however, two types of randomness affect this process. First, the exact time of generation may slightly deviate from $t_{\mathrm{qb}}/2$ due to physical limitations in the photon-pair generation mechanism. However, this timing uncertainty is significantly lower than the time jitter introduced by the SPAD detectors during avalanche triggering \cite{gobby2004quantum}. Therefore, to avoid excessive notation and complexity, and based on practical averaging over many trials, we neglect the time uncertainty at the source.
Second, the number of entangled pairs generated per interval follows a Poisson distribution due to the probabilistic nature of spontaneous parametric down-conversion (SPDC) or similar quantum generation mechanisms \cite{liu2019solid}. The probability of generating exactly \( n_t \) photon pairs in a given time interval is given by \cite{liu2019solid}:
\begin{align}
	P(n_t) = \frac{{\mu_t^{n_t} e^{-\mu_t}}}{n_t!},
	\label{eq:poisson}
\end{align}
where \( \mu_t \) is the mean number of photon pairs generated per interval.

\subsection{Signal Representation} \label{sec_seq}
As discussed, each entangled photon pair is split such that one photon is directed towards a reference block (serving as a local basis) while the other is reflected and eventually detected at the receiver. Within the reference block, the incoming photon generates two vectors of data. The first vector, denoted by $\mathbf{b}_\mathrm{ref}$, contains the logical bit value (0 or 1), inferred from the photon’s polarization state, e.g., horizontal (H) or vertical (V). The second vector, denoted by $\mathbf{t}_\mathrm{ref}$, holds the precise arrival times of the photons as recorded by the SPAD array.

Similarly, at the receiver, detection of the reflected photon results in the generation of two corresponding vectors. The first, $\mathbf{b}_\mathrm{rx}$, includes the detected bit values based on polarization measurement, and the second, $\mathbf{t}_\mathrm{rx}$, consists of their respective detection times. 
\textcolor{blue}{These four vectors, $\mathbf{b}_{\mathrm{ref}}$, $\mathbf{t}_{\mathrm{ref}}$, $\mathbf{b}-{\mathrm{rx}}$, and $\mathbf{t}_{\mathrm{rx}}$, form the fundamental observables used in our synchronization framework.
Each vector has a length determined by the total acquisition duration $t_{\mathrm{aq}}$ and the photon-pair generation rate $R_{\mathrm{qb}} = 1/t_{\mathrm{qb}}$. Specifically, the total sequence length is given by:}
\begin{align}
	L_{\mathrm{seq}} = {t_{\mathrm{aq}}}/{t_{\mathrm{qb}}},
	\label{eq:L_seq}
\end{align}
which corresponds to the number of time slots used for probing all spatial grid cells.

Since the acquisition process involves sequentially directing the beam toward each of the $N_{\mathrm{gr}}$ spatial grid cells, the measurement vectors are naturally partitioned according to these grid locations. We define sub-vectors:
\begin{align}
	\mathbf{b}_\mathrm{ref}^{(j)}, \quad \mathbf{t}_\mathrm{ref}^{(j)}, \quad
	\mathbf{b}_\mathrm{rx}^{(j)}, \quad \mathbf{t}_\mathrm{rx}^{(j)} \quad \text{for } j = 1, \dots, N_{\mathrm{gr}},
\end{align}
where each sub-vector corresponds to the data collected during the time window in which the Gaussian beam was directed to the $j$th grid cell. Each sub-vector has equal length:
\begin{align}
	L_{\mathrm{sv}} = {L_{\mathrm{seq}}}/{N_{\mathrm{gr}}}.
	\label{eq:L_sv}
\end{align}
For example, at a rate of $R_{\mathrm{qb}} = 1$ Gbps, total acquisition time of $t_{\mathrm{aq}} = 100\ \mu\mathrm{s}$, and $N_{\mathrm{gr}} = 100$ grid points, we obtain $L_{\mathrm{seq}} = 10^5$ and $L_{\mathrm{sv}} = 10^3$ samples per sub-vector. 


\subsubsection{Reference Vectors $\mathbf{b}_\mathrm{ref}^{(j)}$ and $\mathbf{t}_\mathrm{ref}^{(j)}$}
In each acquisition cycle, the quantum transmitter probabilistically generates entangled photon pairs at a rate of $R_\mathrm{qb} = 1 / t_\mathrm{qb}$. The reference unit records data only for time intervals in which exactly one photon pair is successfully generated and delivered, whereas events with zero or multiple generated photon pairs are considered ambiguous and therefore ignored, that is, filtered out.

As a result, the $j'$th element of the polarization bit sub-vector $\mathbf{b}_\mathrm{ref}^{(j)}$ corresponding to the $j$th grid cell can take one of the following three values:
\begin{itemize}
	\item 0: Horizontally polarized photon detected,
	\item 1: Vertically polarized photon detected,
	\item $\varnothing$: Invalid or filtered due to zero or multiple photon pairs in the interval.
\end{itemize}

Each valid bit is associated with a precise timestamp stored in the sub-vector $\mathbf{t}_\mathrm{ref}^{(i)}$. The $j'$th timestamp of the $j$th sub-vector is modeled as
\begin{align} \label{eqe2}
	t_\mathrm{ref}^{(j)}(j') = t_0 + \left((j - 1) L_\mathrm{sv} + j' \right) t_\mathrm{qb} + t_\mathrm{e,ref},
\end{align}
where $t_0$ denotes the start time of the acquisition, $t_\mathrm{qb}$ is the duration of each qubit interval, and $t_\mathrm{e,ref} \sim \mathcal{N}(0, \sigma_\mathrm{spad}^2)$ represents Gaussian-distributed temporal jitter introduced by the SPAD detector electronics.

\subsubsection{Receiver Vectors $\mathbf{b}_\mathrm{rx}^{(j)}$ and $\mathbf{t}_\mathrm{rx}^{(j)}$}
Unlike the reference unit, where photons are directly routed and almost deterministically received (conditional on valid generation), the receiver is affected by quantum channel losses and stochastic photon detection. In each time slot $j'$ of the $j$th acquisition window, the probability of successfully detecting a reflected photon follows a Poisson process with mean $\mu_{\mathrm{ch},j} \ll 1$, i.e.,
\begin{align}  \label{eqg3}
	P_\mathrm{rx}(n_{rx}) = \frac{\mu_{\mathrm{ch},j}^{n_{rx}} e^{-\mu_{\mathrm{ch},j}}}{n_{rx}!}, \quad n_{rx} = 0, 1, 2, \dots
\end{align}
This results in a sparse detection process, where most time intervals yield no photon detection.

To ensure a consistent comparison with the reference side, we impose a filtering condition: the receiver vector $\mathbf{b}_\mathrm{rx}^{(j)}(j')$ is defined \emph{only} if the corresponding reference bit $\mathbf{b}_\mathrm{ref}^{(j)}(j')$ is valid (i.e., $\mathbf{b}_\mathrm{ref}^{(j)}(j') \in \{0,1\}$), and if at least one photon is detected at the receiver in the corresponding interval. Otherwise, we assign $\mathbf{b}_\mathrm{rx}^{(j)}(j') = \varnothing$.

Each valid detection is also associated with a timestamp modeled as
\begin{align}  \label{eqe1}
	t_\mathrm{rx}^{(j)}(j') = t_0 + \left((j - 1) L_\mathrm{sv} + j' \right) t_\mathrm{qb} + t_\mathrm{ch} + t_\mathrm{e,rx},
\end{align}
where $t_\mathrm{ch}$ denotes the round-trip propagation delay through the quantum free-space channel, and $t_\mathrm{e,rx} \sim \mathcal{N}(0, \sigma_\mathrm{spad}^2)$ captures the timing uncertainty induced by the SPAD detector at the receiver.

\section{Channel Modeling}
In each acquisition time slot \( t_j \), as previously described, the entangled photon transmitter directs a Gaussian beam towards the center of a selected grid cell indexed by \( (j_x, j_y) \), with spatial coordinates denoted by \( \mathbf{p}(j_x, j_y) \). 

\begin{remark}
	Note that in each time slot of duration $t_{\mathrm{qb}}$, only a single photon is transmitted. This photon may be captured by one of the CCR units with a certain probability, or it may miss all of them and remain undetected.
\end{remark}

In the round-trip photon propagation process, the probability that a single transmitted photon is ultimately received at the ground station via the $i$th CCR unit depends on several sequential factors. These include the probability that the photon spatially falls onto the $i$th CCR unit, the likelihood that the reflected photon propagates back toward the receiver and lands within its aperture, and the transmission efficiencies associated with atmospheric and aperture-related losses along both the forward and return paths.
Accordingly, the overall single-photon reception probability via the $i$th CCR unit and the $j$th beam direction (corresponding to the $j$th grid cell) is given by
\begin{align}
	P_{\mathrm{rec},i,j} = \underbrace{P_{\mathrm{hit},i,j} \cdot P_{\mathrm{ap},i,j}}_{P_{\mathrm{hap},i,j}}
	 \cdot h_{\mathrm{a},i,j} \cdot h_{\mathrm{L,a}} \cdot h_{\mathrm{L,c}},
	\label{eq:photon_reception_i}
\end{align}
where $i \in \{1, \dots, N_{\mathrm{ar}}\}$ and $j \in \{1, \dots, N_{\mathrm{gr}}\}$, and
\begin{itemize}
	\item $P_{\mathrm{hit},i,j}$ is the probability that the transmitted photon impinges on the $i$th CCR unit when the Gaussian beam is directed toward the $j$th grid cell,
	\item $P_{\mathrm{ap},i,j}$ is the probability that the photon reflected from the $i$th CCR unit is successfully captured by the receiver’s aperture along the return path,
	\item $h_{\mathrm{a},i,j}$ is the atmospheric turbulence coefficient for the round-trip path,
	\item $h_{\mathrm{L,a}}$ is the atmospheric attenuation factor for the round-trip path,
	\item $h_{\mathrm{L,c}}$ is the optical coupling loss factor at the transmitter and receiver apertures.
\end{itemize}

\begin{proposition}
Under the assumption that all CCR units have identical aperture area \( A_{\mathrm{ar},i} = A_{\mathrm{ar}} \), the combined probability that a photon both hits the $i$th CCR and is successfully captured after reflection is given by
\begin{align}
	P_{\mathrm{hap},i,j} 
	&= \frac{2 A_{\mathrm{ar}}}{\pi w_z^2} 
		\exp\left( -\frac{2 \| \mathbf{p}_{\mathrm{ar},i} - \mathbf{p}(j_x, j_y) \|^2}{w_z^2} \right) \nonumber \\
	& ~~~\times \left( 1 - \exp\left( -\frac{2 A_{\mathrm{ar}} r_{\mathrm{ap}}^2}{\lambda^2 L_{\mathrm{tar}}^2} \right) \right),
	\label{eqg2}
\end{align}
\textcolor{blue}{where \( w_z \) is the beam waist of the Gaussian beam at the target plane.}
\end{proposition}

\begin{IEEEproof}
	Please refer to Appendix \ref{AppA}.
\end{IEEEproof}
This result provides a closed-form and spatially-resolved expression for the probability that a photon is successfully reflected and detected, capturing the effect of both beam spread and geometric layout of the CCR array. 

The remaining multiplicative factors in \eqref{eq:photon_reception_i} account for atmospheric and coupling losses.
\( h_{\mathrm{a},i,j} \) represents the fading coefficient due to atmospheric turbulence on the round-trip path. We adopt the widely used Gamma-Gamma model, where the probability density function (PDF) of \( h_{\mathrm{a},i,j} \) is given by \cite{andrews2005laser}:
\begin{align}
	f_{h_{\mathrm{a},i,j}}(h) = \frac{2 (\alpha \beta)^{\frac{\alpha + \beta}{2}}}{\Gamma(\alpha) \Gamma(\beta)} h^{\frac{\alpha + \beta}{2} - 1} K_{\alpha - \beta} \left( 2 \sqrt{\alpha \beta h} \right),
\end{align}
where \( \alpha \) and \( \beta \) are the turbulence parameters, and \( K_{\nu}(\cdot) \) is the modified Bessel function of the second kind. \( h_{\mathrm{L,a}} \) models the deterministic atmospheric attenuation (e.g., due to scattering and absorption) along the round-trip path and is expressed as:
\begin{align}
	h_{\mathrm{L,a}} = \exp(-\sigma_{\mathrm{atm}} \cdot 2L_{\mathrm{tar}}),
\end{align}
	where \( \sigma_{\mathrm{atm}} \) is the atmospheric extinction coefficient and \( L_{\mathrm{tar}} \) is the target distance.
\( h_{\mathrm{L,c}} \) denotes the geometric optical coupling efficiency at both the transmitter and receiver apertures and is typically modeled as a constant depending on alignment and aperture matching.

For a given $j=(j_x, j_y)$, the overall single-photon reception probability in each time slot, denoted by $P_{\mathrm{rec},j}$, is obtained by aggregating the contributions from all $N_{\mathrm{ar}}$ CCR units. Using the result of Proposition~1 in \eqref{eqg2} and applying the grid cell decomposition from \eqref{eq:gridcell_position}, the total reception probability becomes:
\begin{align}
	&P_{\mathrm{rec},j} 
	=  \frac{2 A_{\mathrm{ar}}}{\pi w_z^2}  h_{\mathrm{L,a}}  h_{\mathrm{L,c}}
	\left( 1 - \exp\left( -\frac{2 A_{\mathrm{ar}} r_{\mathrm{ap}}^2}{\lambda^2 L_{\mathrm{tar}}^2} \right) \right) \nonumber \\
	&\resizebox{0.88\hsize}{!}{$\times \sum_{i=1}^{N_{\mathrm{ar}}} h_{\mathrm{a},i,j} \cdot  
	\exp\left( -\frac{2 \| \mathbf{p}_{\mathrm{ar},i} - \mathbf{p}'(j_x, j_y) - \mathbf{r}_{\mathrm{dev}} \|^2}{w_z^2} \right). $} 
	\label{eq:Precj_updated}
\end{align}

\begin{remark}
	Although $P_{\mathrm{rec},j}$ in \eqref{eq:Precj_updated} represents a single-photon reception probability, it is itself a random variable. This is because it depends on $N_{\mathrm{ar}}$ distinct realizations of the turbulence-induced fading coefficients $h_{\mathrm{a},i,j}$ and the unknown spatial offset vector $\mathbf{r}_{\mathrm{dev}}$. Consequently, the reception probability fluctuates across different realizations of atmospheric turbulence and target offset.
\end{remark}

\begin{proposition} \label{prop:prec_gaussian_final}
	The PDF of $P_{\mathrm{rec},j}$ conditioned on $\mathbf{r}_{\mathrm{dev}}$ is derived as:
	\begin{align}
		&f_{P_{\mathrm{rec},j} | \mathbf{r}_{\mathrm{dev}}}(p) 
		= \frac{1}{\sqrt{2\pi c_0^2 c_{\alpha,\beta} \sum_{i=1}^{N_{\mathrm{ar}}} w_{i,j,\mathbf{r}_{\mathrm{dev}}}^2 }} \nonumber \\
		&~~~\times \exp\left( -\frac{(p - c_0 \cdot \sum_{i=1}^{N_{\mathrm{ar}}} w_{i,j,\mathbf{r}_{\mathrm{dev}}})^2}
		{2c_0^2 c_{\alpha,\beta} \sum_{i=1}^{N_{\mathrm{ar}}} w_{i,j,\mathbf{r}_{\mathrm{dev}}}^2} \right),
		\label{eq:precj_pdf}
	\end{align}
	where 
	\begin{align}
		\left\{
		\begin{aligned}
			& w_{i,j,\mathbf{r}_{\mathrm{dev}}} \triangleq \exp\left( -\frac{2 \| \mathbf{p}_{\mathrm{ar},i} - \mathbf{p}'(j_x, j_y) - \mathbf{r}_{\mathrm{dev}} \|^2}{w_z^2} \right), \\
			& c_0 = \frac{2 A_{\mathrm{ar}}}{\pi w_z^2}  h_{\mathrm{L,a}}  h_{\mathrm{L,c}}
			\left( 1 - \exp\left( -\frac{2 A_{\mathrm{ar}} r_{\mathrm{ap}}^2}{\lambda^2 L_{\mathrm{tar}}^2} \right) \right), \\
			&c_{\alpha,\beta} = \left( \frac{1}{\alpha} + \frac{1}{\beta} + \frac{1}{\alpha\beta} \right)
		\end{aligned}
		\right. \label{eq:mean_var_h}
	\end{align}
	
\end{proposition}

\begin{IEEEproof}
	Please refer to Appendix~\ref{AppB}.
\end{IEEEproof}

Proposition~\ref{prop:prec_gaussian_final} provides the conditional distribution of $P_{\mathrm{rec},j}$ given $\mathbf{r}_{\mathrm{dev}}$, capturing the statistical behavior of single-photon reception under a fixed spatial offset. To obtain the unconditional distribution, one must marginalize over $\mathbf{r}_{\mathrm{dev}}$:
\begin{align} \label{eqg8}
	f_{P_{\mathrm{rec},j}}(p) = \int f_{P_{\mathrm{rec},j} | \mathbf{r}_{\mathrm{dev}}}(p) \cdot f_{\mathbf{r}_{\mathrm{dev}}}(\mathbf{r}) ~d\mathbf{r}.
\end{align}
However, due to the nonlinear dependence of the spatial weights $w_{i,j,\mathbf{r}_{\mathrm{dev}}}$ on $\mathbf{r}_{\mathrm{dev}}$, as seen in \eqref{eq:mean_var_h}, this integral lacks a closed-form and becomes computationally intractable for general fading configurations. Nevertheless, Proposition~\ref{prop:prec_gaussian_final} effectively reduces a high-dimensional marginalization over $N_{\mathrm{ar}}$ independent fading coefficients to a two-dimensional integration over $\mathbf{r}_{\mathrm{dev}}$, significantly simplifying numerical evaluation while closely matching simulation results.

Finally, each received photon is routed to one of the SPAD detectors based on its polarization, and the probability of detection is scaled by the detector efficiency $\eta_{\mathrm{spad}}$ \cite{dabiri2025unified}. Accordingly, the expected number of received photons per time slot is given by:
\begin{align}  \label{eqg6}
	\mu_{\mathrm{ch},j} = \eta_{\mathrm{spad}} \cdot P_{\mathrm{rec},j},
\end{align}
and its conditional distribution given $\mathbf{r}_{\mathrm{dev}}$ is \cite{shynk2012probability}:
\begin{align} \label{eqg7}
	f_{\mu_{\mathrm{ch},j} | \mathbf{r}_{\mathrm{dev}}}(\mu) 
	= \frac{1}{\eta_{\mathrm{spad}}} \cdot f_{P_{\mathrm{rec},j} | \mathbf{r}_{\mathrm{dev}}} \left( \frac{\mu}{\eta_{\mathrm{spad}}} \right),
\end{align}

\begin{remark}
	Although $\mu_{\mathrm{ch},j}$ is itself a random variable, it serves as the deterministic rate parameter in the Poisson distribution of received photon counts at the SPAD detectors, as defined in \eqref{eqg3}. Therefore, under any instantaneous channel realization, the detected photon count $n_{\mathrm{rx}}$ in a time slot follows a Poisson distribution with a mean that depends on the specific atmospheric and spatial configuration.
\end{remark}

In addition to signal photons, background photons from ambient light, especially under daylight conditions, may also reach the receiver. These photons follow a Poisson distribution with mean count $\mu_{\mathrm{bg}}$ per time slot. 
A more critical challenge posed by background photons is their temporal randomness. As previously discussed, the transmitter emits each signal photon around the midpoint of the qubit slot, i.e., at $t = t_{\mathrm{qb}}/2$, to facilitate precise temporal alignment. In contrast, background photons arrive uniformly at random throughout the entire qubit time slot $[0, t_{\mathrm{qb}}]$. This timing uncertainty can significantly degrade synchronization precision and affect qubit discrimination. The probability density function of the arrival time $t_{\mathrm{bg}}$ of a background photon is:
\begin{align}  \label{eqg4}
	f_{t_{\mathrm{bg}}}(t) = 
	\begin{cases}
		\frac{1}{t_{\mathrm{qb}}}, & -t_{\mathrm{qb}}/2 \leq t \leq t_{\mathrm{qb}}/2, \\
		0, & \text{otherwise}.
	\end{cases}
\end{align}

Polarization-flip errors may occur due to imperfections in the optical path, causing an $H$-polarized photon to be detected as $V$, or vice versa. However, since only unscattered photons reflected by CCRs are likely to reach the receiver, and high-quality optics minimize such effects, the actual error is typically small. We conservatively assume a worst-case polarization error of $\mathbb{P}_{\mathrm{pol}} = 0.1$.

\section{Quantum Synchronization}
To achieve wireless quantum synchronization, we need to estimate and correct temporal offsets between the transmitter and the target using entangled photon pairs. We evaluate the synchronization accuracy under practical conditions, including turbulence, detection noise, and background interference.
The first essential step in synchronization is to match the received quantum bit sequence $\mathbf{b}_\mathrm{rx} = [\mathbf{b}_\mathrm{rx}^{(1)}, \dots, \mathbf{b}_\mathrm{rx}^{(N_{\mathrm{gr}})}]$ to the reference sequence $\mathbf{b}_\mathrm{ref} = [\mathbf{b}_\mathrm{ref}^{(1)}, \dots, \mathbf{b}_\mathrm{ref}^{(N_{\mathrm{gr}})}]$, both of which were defined in Section~\ref{sec_seq}. 
Due to the probabilistic nature of entangled photon generation and the inherent channel losses, the average number of detected photons per slot, denoted by $\mu_{\mathrm{ch}}$, typically satisfies $\mu_{\mathrm{ch}} \ll 1$. As a result, the quantum channel is extremely lossy, and the majority of entries in the received bit sequence $\mathbf{b}_\mathrm{rx}$ are null.
In addition, the round-trip propagation introduces a synchronization offset that must be accounted for during sequence alignment. Assuming an initial position uncertainty of up to 1 meter, a photon-pair generation rate of 1~GHz, and a light speed of $3 \times 10^8$~m/s, the maximum timing misalignment in the received sequence $\mathbf{b}_\mathrm{rx}$ caused by round-trip delay uncertainty is approximately
\[
\Delta N_{\max} = \left\lceil \frac{2 \times 1~\mathrm{m}}{3 \times 10^8~\mathrm{m/s}} \cdot 10^9~\mathrm{Hz} \right\rceil = 7~\text{slots},
\]
where the factor of 2 accounts for the two-way propagation. 

Achieving ps-level synchronization requires a sufficient number of valid photon detections in the received bit sequence. Let $N_{\mathrm{tot}}$ denote the total number of non-null entries in $\mathbf{b}_\mathrm{rx}$. A larger $N_{\mathrm{tot}}$ improves statistical averaging and temporal resolution, which is essential for high-precision alignment \cite{lyu2023nanomotion}. When the synchronization window is sufficiently long, robust techniques such as cross-correlation \cite{gobby2004quantum} or maximum-likelihood matching \cite{liao2017satellite} can reliably align $\mathbf{b}_\mathrm{rx}$ with $\mathbf{b}_\mathrm{ref}$. 
Given that this step is not a dominant bottleneck and well-established solutions exist, we assume for the remainder of this paper that the received sequence has been successfully aligned with the reference, and focus instead on modeling and analyzing the synchronization accuracy.

We now focus on the analysis of the received timestamp vector $\mathbf{t}_\mathrm{rx}$, where each element $t_\mathrm{rx}^{(j)}(j')$ was previously defined in \eqref{eqe1}. Once the correct alignment index is identified through matching, the composite time index term $\left((j - 1) L_\mathrm{sv} + j' \right) t_\mathrm{qb}$ becomes known for each valid detection.
By subtracting this deterministic offset from \eqref{eqe1}, the timestamp of each received photon can be rewritten as:
\begin{align} \label{eq:trx_aligned}
	\tilde{t}_\mathrm{rx}^{(j)}(j')  = 
	\begin{cases}
		t_0 + t_\mathrm{ch} + t_{\mathrm{e,rx},j'}, & \text{if } s_{j'} = 0,\\
		t_0 + t_\mathrm{ch} + t_{\mathrm{bg},j'}, & \text{if } s_{j'} = 1,
	\end{cases}
\end{align}
where $s_{j'} = 0$ indicates that the recorded signal is due to a reflected photon, and $s_{j'} = 1$ indicates that it is due to a background photon. Note that $s_{j'}$ values are not known in practice and are introduced solely for synchronization error analysis.
Similarly, each reference timestamp in \eqref{eqe2} can be simplified by removing the deterministic time index term, yielding:
\begin{align}
	\tilde{t}_\mathrm{ref}^{(j)}(j') = t_0 + t_\mathrm{e,ref,j'}.
	\label{eq:ref_aligned}
\end{align}
By subtracting the aligned reference timestamps from the aligned received timestamps and averaging over all valid detection slots, the estimated synchronization delay $\hat{t}_{\mathrm{ch}}$ is given by:
\begin{align} \label{eq:tch_estimation_final}
	\hat{t}_\mathrm{ch} = t_\mathrm{ch} + n_\mathrm{ch},
\end{align}
where the estimation noise $n_\mathrm{ch}$ is given by:
\begin{align}
	n_\mathrm{ch} &= \frac{1}{N_{\mathrm{tot}}} \left( \sum_{j'=1}^{N_{\mathrm{tot}}} (t_{\mathrm{e,rx},j'} - t_{\mathrm{e,ref},j'}) \cdot (1 - s_{j'})   \right. \nonumber \\
	&~~~ \left.+  \sum_{j'=1}^{N_{\mathrm{tot}}} (t_{\mathrm{bg},j'} - t_{\mathrm{e,ref},j'}) \cdot s_{j'}  \right),
	\label{eq:nch_model}
\end{align}
with $ N_{\mathrm{tot}} = N_{\mathrm{sig}} + N_{\mathrm{bg}} $, \(N_{\mathrm{sig}} = \sum_{j'=1}^{N_{\mathrm{tot}}} (1 - s_{j'})\) and \(N_{\mathrm{bg}} = \sum_{j'=1}^{N_{\mathrm{tot}}} s_{j'}\) denoting the number of signal and background detections, respectively.

The random variable \(n_\mathrm{ch}\) reflects the combined effect of detector jitter and background photon randomness, scaled by the signal and background counts. Notably, \(N_{\mathrm{sig}}\) is itself a random variable, determined by the probabilistic nature of photon detection and influenced by key channel parameters such as \(\mu_{\mathrm{ch}}\), \(t_{\mathrm{aq}}\), and \(R_{\mathrm{qb}}\). As a result, the synchronization noise inherently depends on both measurement noise and the stochastic structure of the channel.

As observed from \eqref{eq:nch_model}, all noise components involved in the synchronization process, namely the SPAD induced jitter terms $t_{\mathrm{e,rx}}$ and $t_{\mathrm{e,ref}}$ as well as the background arrival time $t_{\mathrm{bg}}$, are modeled as zero mean random variables. Consequently, since $n_{\mathrm{ch}}$ is a linear combination of these zero-mean terms, it also has zero mean $\mathbb{E}[n_{\mathrm{ch}}] = 0$.
To better characterize the impact of these uncertainties on synchronization performance, it is more insightful to examine the variance of $n_{\mathrm{ch}}$. This variance captures the stochastic fluctuations introduced by both measurement noise and background randomness, and is derived analytically in Proposition~\ref{prop:voise_var}.

\begin{proposition} \label{prop:voise_var}
	The expected variance of the quantum synchronization error $n_{\mathrm{ch}}$ over all realizations of $N_{\mathrm{sig}}$ and $N_{\mathrm{bg}}$ is given by:
	\begin{align} \label{eq:var_nch_final2}
		&\mathrm{Var}(n_\mathrm{ch})  
		= \sum_{N_{\mathrm{sig}}=N_{\mathrm{t,min}}}^{N_{\mathrm{s,max}}}
		\sum_{N_{\mathrm{bg}} = \max(0, N_{\mathrm{t,min}} - N_{\mathrm{sig}})}^{N_{\mathrm{b,max}}}   \\
		&\left[
		\frac{2 N_{\mathrm{sig}} \sigma_{\mathrm{spad}}^2 
			+ N_{\mathrm{bg}} \left( \frac{t_{\mathrm{qb}}^2}{12} + \sigma_{\mathrm{spad}}^2 \right)}
		{(N_{\mathrm{sig}} + N_{\mathrm{bg}})^2}
		\right]  \mathbb{P}(N_{\mathrm{sig}}) \cdot \mathbb{P}(N_{\mathrm{bg}}), \nonumber
	\end{align}
	where $N_{\mathrm{t,min}}$ is the minimum required number of valid (non-null) photon detections needed to perform meaningful synchronization, $N_{\mathrm{s,max}}$ and $N_{\mathrm{b,max}}$ denote upper bounds on signal and background counts based on system and channel parameters, and $\mathbb{P}(N_{\mathrm{sig}})$ and $\mathbb{P}(N_{\mathrm{bg}})$ are the probability mass functions of signal and background photon counts, respectively.
\end{proposition}

\begin{IEEEproof}
	Please refer to Appendix~\ref{AppC}.
\end{IEEEproof}

Proposition~\ref{prop:voise_var} provides an expression for the variance of the synchronization error $n_{\mathrm{ch}}$ in terms of SPAD timing jitter $\sigma_{\mathrm{spad}}$, the qubit slot duration $t_{\mathrm{qb}}$, the relative background ratio $N_{\mathrm{bg}}$, and received quantum photons $N_{\mathrm{sig}}$.

Moreover, $N_{\mathrm{t,min}}$ defines the minimum total number of valid (non-null) photon detections required for successful synchronization. If the total number $N_{\mathrm{tot}} = N_{\mathrm{sig}} + N_{\mathrm{bg}}$ falls below this threshold, the system enters synchronization outage and cannot reliably perform sequence matching. Therefore, $\mathbb{P}_\mathrm{out}$ quantifies the probability that the system fails to perform synchronization due to insufficient valid detections, and thus serves as a key performance metric in evaluating synchronization robustness:
\begin{align}
	\label{P_out1}
	\mathbb{P}_\mathrm{out}
	= \sum_{N_{\mathrm{sig}}=0}^{N_{\mathrm{t,min}}} 
	\sum_{N_{\mathrm{bg}}=0}^{\max(0, N_{\mathrm{t,min}} - N_{\mathrm{sig}})}  
	\mathbb{P}(N_{\mathrm{sig}}) \mathbb{P}(N_{\mathrm{bg}}).
\end{align}
Note that threshold $N_{\mathrm{t,min}}$ is selected to satisfy two conditions: (i) the number of detected signal photons must exceed a minimum value $N_{\mathrm{s,min}}$ to enable reliable matching with the reference sequence; and (ii) the probability of signal dominance must be high, i.e., $N_{\mathrm{sig}} \gg N_{\mathrm{bg}}$. To ensure this, we define:
\begin{align}
	N_{\mathrm{t,min}} = \max\left(N_{\mathrm{s,min}},~ m \cdot \mu_{\mathrm{bg}}\right),
\end{align}
where $m$ is a design parameter controlling the background suppression margin.

Both the synchronization error variance and the outage probability are fundamentally governed by the distribution of detected signal photons, $\mathbb{P}(N_{\mathrm{sig}})$, which is characterized in Proposition~\ref{prop:N_sig}.

\begin{proposition} \label{prop:N_sig}
	The pdf of $N_{\mathrm{sig}}$  is derived as:
	\begin{align} \label{eq:Nsig_poisson_r_div}
		\mathbb{P}(N_{\mathrm{sig}} = n) = \int \frac{\mu_{\mathrm{ch}}^n\exp\left(-\mu_{\mathrm{ch}} \right)      }{n!} 
		f_{\mu_{\mathrm{ch}}}(\mu) d\mu	,
	\end{align} 
	\begin{align}
		\label{f_mu}
		f_{\mu_{\mathrm{ch}}}(\mu) = \int f_{\mu_{\mathrm{ch}} | \mathbf{r}_{\mathrm{dev}}}(\mu) 
		f_{\mathbf{r}_{\mathrm{dev}}}(r)~dr,
	\end{align}
	\begin{align}
		&f_{\mu_{\mathrm{ch}} | \mathbf{r}_{\mathrm{dev}}}(\mu) = 
		\frac{1}{\sqrt{2\pi \eta_{\mathrm{spad}}^2 \lambda_t^2 c_0^2 c_{\alpha,\beta} \sum_{j=1}^{N_{\mathrm{gr}}} \sum_{i=1}^{N_{\mathrm{ar}}} w_{i,j,\mathbf{r}_{\mathrm{dev}}}^2 }} \nonumber \\
		& \times \exp\left( -\frac{\left( \mu - \eta_{\mathrm{spad}} \lambda_t c_0 \sum_{j=1}^{N_{\mathrm{gr}}} \sum_{i=1}^{N_{\mathrm{ar}}} w_{i,j,\mathbf{r}_{\mathrm{dev}}} \right)^2}
		{2 \eta_{\mathrm{spad}}^2 \lambda_t^2 c_0^2 c_{\alpha,\beta} \sum_{j=1}^{N_{\mathrm{gr}}} \sum_{i=1}^{N_{\mathrm{ar}}} w_{i,j,\mathbf{r}_{\mathrm{dev}}}^2} \right),
		\label{eq:mu_ch_prime_pdf_final2}
	\end{align}
	and $\mu_{\mathrm{ch}} = \lambda_t
	\sum_{j=1}^{N_{\mathrm{gr}}} \mu_{\mathrm{ch},j}$, $\lambda_t = L_{\mathrm{sv}} \mu_t e^{-\mu_t}$, and the parameters $w_{i,j,\mathbf{r}_{\mathrm{dev}}}$, $c_0$, and $c_{\alpha,\beta}$ are provided in \eqref{eq:mean_var_h}.	
\end{proposition}

\begin{IEEEproof}
	Please refer to Appendix~\ref{AppD}.
\end{IEEEproof}

Although the expression in Proposition~\ref{prop:N_sig} involves a double integration over the spatial offset, we adopt a worst-case approximation that significantly reduces complexity by assuming radial displacement along a single axis. Under this assumption, the distribution $f_{\mu_{\mathrm{ch}}}(\mu)$ simplifies to a one-dimensional integral with respect to a Rayleigh-distributed deviation. 
This formulation is computationally efficient, taking less than one second to evaluate in MATLAB, and as demonstrated in Section~\ref{sec:sim}, it provides an accurate and conservative approximation of the synchronization system under practical misalignment conditions.

To model the background photon count $N_{\mathrm{bg}}$, as discussed in Appendix \ref{AppC}, we note that each of the $\lambda_t = N_{\mathrm{gr}} L_{\mathrm{sv}} \mu_t e^{-\mu_t}$ valid time slots may independently register a background detection. A detection occurs if (i) exactly one background photon arrives in the slot, following a Poisson process with rate $\mu_{\mathrm{bg}}$, and (ii) no signal photon is present. 
The probability of observing exactly one background photon in a slot is $\mu_{\mathrm{bg}} e^{-\mu_{\mathrm{bg}}}$. Given the highly lossy nature of the quantum channel, the joint occurrence of signal and background photons in the same slot is negligible, and can be safely ignored for analytical purposes.
Therefore, the total number of background photon detections $N_{\mathrm{bg}}$ follows a Poisson distribution with mean:
\begin{align}
	\mathbb{E}[N_{\mathrm{bg}}] = \lambda_t \cdot \mu_{\mathrm{bg}} e^{-\mu_{\mathrm{bg}}}.
\end{align}
Hence, its probability mass function is given by:
\begin{align}
	\mathbb{P}(N_{\mathrm{bg}} = n) = 
	\frac{ \left( \lambda_t \mu_{\mathrm{bg}} e^{-\mu_{\mathrm{bg}}} \right)^n }{n!} 
	e^{ - \lambda_t \mu_{\mathrm{bg}} e^{-\mu_{\mathrm{bg}}} }.
\end{align}

It is important to note that the background photon arrival rate $\mu_{\mathrm{bg}}$ is not constant and may vary significantly over time, especially in outdoor environments. Its fluctuations are primarily influenced by environmental lighting conditions and the relative position of the Sun. These variations can affect the number of false detections and thus impact synchronization performance. In the following section, we evaluate the impact of different background levels on synchronization accuracy under representative daytime and nighttime scenarios.

\begin{figure*}
	\centering
	\subfloat[] {\includegraphics[width=1.7 in]{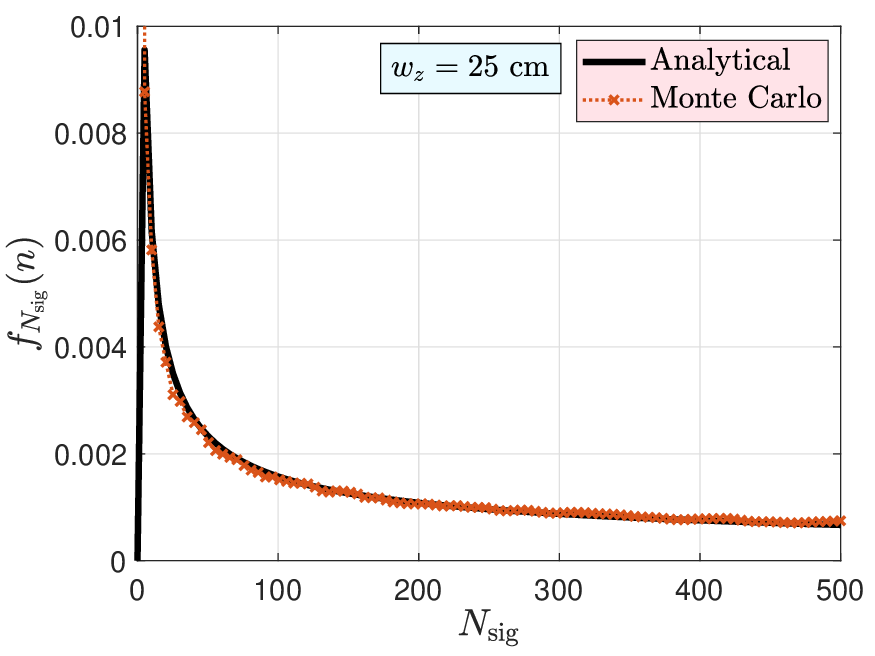}
		\label{cz1}
	}
	\hfill
	\subfloat[] {\includegraphics[width=1.7 in]{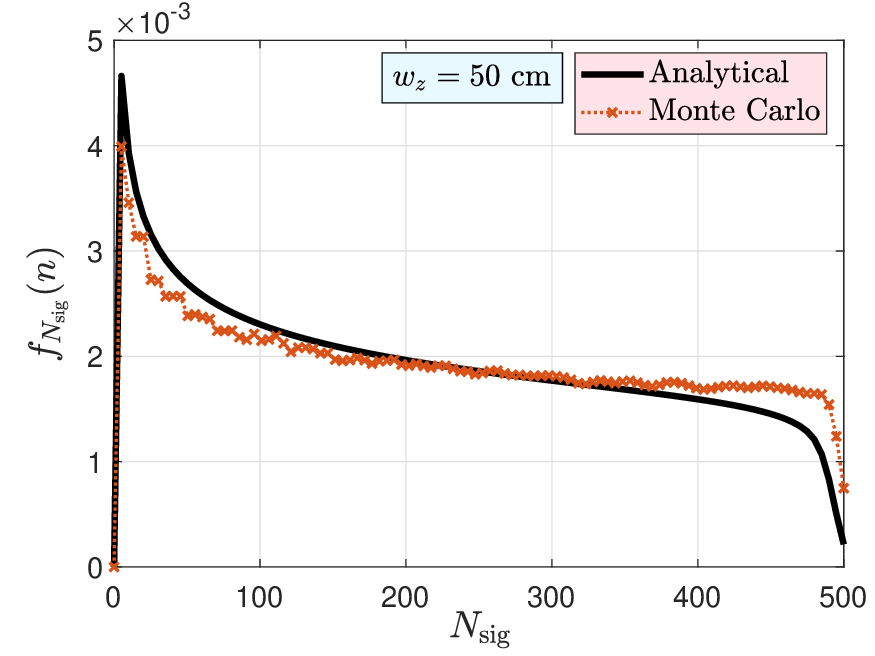}
		\label{cz2}
	}
    \hfill
    \subfloat[] {\includegraphics[width=1.7 in]{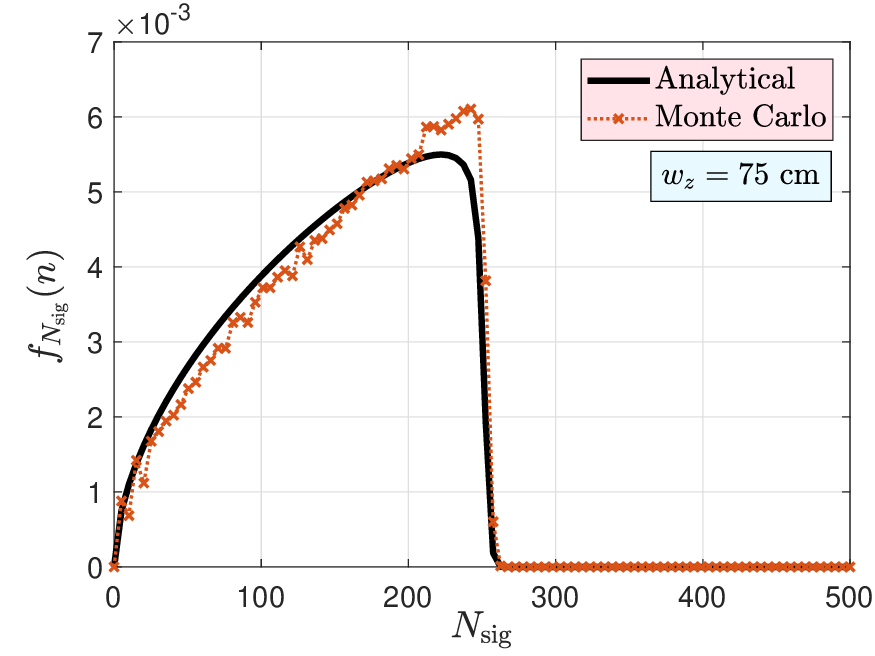}
    	\label{cz3}
    }
    \hfill
    \subfloat[] {\includegraphics[width=1.7 in]{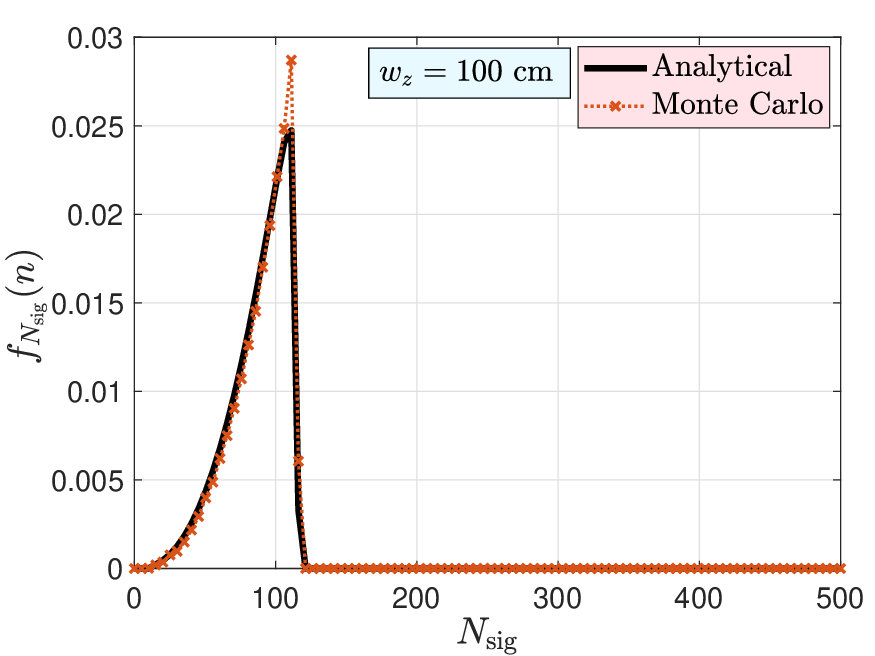}
    	\label{cz4}
    }
	\caption{Analytical and simulation results for the distribution \( \mathbb{P}(N_{\mathrm{sig}}) \) under different beam waist values: 
		(a)~\( w_z = 25\,\mathrm{cm} \), 
		(b)~\( w_z = 50\,\mathrm{cm} \), 
		(c)~\( w_z = 75\,\mathrm{cm} \), 
		(d)~\( w_z = 100\,\mathrm{cm} \). 
		Analytical model from Proposition~\ref{prop:N_sig} is validated via Monte Carlo simulations.}
	\label{cz}
\end{figure*}

\begin{table}[!t]
	\centering
	\caption{Default Parameters Used in Simulations}
	\label{tab:sim_params}
	\begin{tabular}{ll|ll}
		\hline
		\textbf{Notation} & \textbf{Value} & \textbf{Notation} & \textbf{Value} \\
		\hline
		$N_{\mathrm{ar}x}$        & 8               & $N_{\mathrm{ar}y}$        & 8 \\
		$N_{\mathrm{ar}}$         & 64              & $d_{\mathrm{ar}}$         & 0.04\,cm \\
		$N_{x}$                   & 10              & $N_{y}$                   & 10 \\
		$d_{\mathrm{gr}}$         & 0.04\,cm        & $\lambda$                 & 1550\,nm \\
		$L_{\mathrm{tar}}$        & 500\,m          & $A_{\mathrm{ar}}$         & 0.0003\,m$^2$ \\
		$r_{\mathrm{ap}}$         & 0.05\,m         & $w_z$                     & 25--125\,cm \\
		$\eta_{\mathrm{spad}}$    & 0.6             & $h_{\mathrm{L,a}}$        & 0.7 \\
		$h_{\mathrm{L,c}}$        & $0.8 \eta_{\mathrm{spad}}$ & $\alpha$                 & 3 \\
		$\beta$                  & 2               & $\sigma_p$              & 10--100\,cm \\
		$\mu_t$                  & 0.5               & $t_{\mathrm{aq}}$         & 100\,$\mu$s \\
		$t_j$                    & 1\,$\mu$s       & $t_{\mathrm{qb}}$         & 1\,ns \\
		$R_{\mathrm{qb}}$         & 2\,Gbps         & $\sigma_{\mathrm{spad}}$  & 50\,ps \\
		$m$                      & 3               & $\mu_{\mathrm{qb}}$       & $10^{-4}$ \\
		\hline
	\end{tabular}
\end{table}

\section{Simulation Results and Discussions} \label{sec:sim}
To validate the analytical derivations and investigate the performance of wireless quantum synchronization under practical conditions, we perform detailed Monte Carlo simulations. 

To estimate the photon detection distribution \( \mathbb{P}(N_{\mathrm{sig}}) \) via Monte Carlo simulation, we first generate \( 10^4 \) independent realizations of the spatial offset vector \( \mathbf{r}_{\mathrm{dev}} = (x_{\mathrm{dev}}, y_{\mathrm{dev}}) \). For each realization of \( \mathbf{r}_{\mathrm{dev}} \), and for every grid cell \( j \in \{1, \dots, N_{\mathrm{gr}}\} \), we compute the reception probability \( P_{\mathrm{rec},j} \) as in \eqref{eq:Precj_updated}. Each \( P_{\mathrm{rec},j} \) involves simulating \( N_{\mathrm{ar}} \) independent CCR paths, and for each path, generating \( 5 \times 10^5 \) random samples of the fading coefficient based on the Gamma-Gamma model. The per-grid reception probabilities \( P_{\mathrm{rec},j} \)s are then aggregated to compute the photon arrival rate, from which \( 5 \times 10^5 \) samples of the signal photon count vector are generated. A histogram of these samples yields the conditional distribution \( \mathbb{P}(N_{\mathrm{sig}} \mid \mathbf{r}_{\mathrm{dev}}) \). Repeating this process over all \( 10^4 \) realizations of \( \mathbf{r}_{\mathrm{dev}} \) and averaging the results provides the final distribution \( \mathbb{P}(N_{\mathrm{sig}}) \) via Monte Carlo estimation. Minor fluctuations observed in the simulation results are attributed to the finite set of \( 10^4 \) randomly generated \( \mathbf{r}_{\mathrm{dev}} \) realizations. Increasing this number would further smooth the distributions, but at the cost of significantly higher computational complexity and runtime.


The default values of the simulation parameters are summarized in Table~\ref{tab:sim_params}. These parameters are not only theoretical constructs, but can be directly mapped to representative, commercially attainable hardware, confirming that the assumed operating regime is achievable with current FSO and single-photon detection technology at \mbox{1550\,nm}. For instance, a beam waist $w_z$ of 0.25--1.25\,m at the receiver is determined by the beam divergence set at the transmitter, which can be adaptively tuned (e.g., using motorized beam expanders or adaptive collimators) to match the link length and maintain the desired spot size under varying atmospheric conditions.
For instance, a beam waist $w_z$ of 0.25--1.25\,m at the receiver results from transmitter beam design (specifically the initial waist size, divergence angle, and link length) with larger $w_z$ required for longer links to limit diffraction, and smaller $w_z$ preferable for shorter links to increase photon density. Divergence can be tuned via beam expanders or adaptive collimators at the transmitter to match atmospheric conditions and target geometry.
Likewise, SPAD efficiency $\eta_{\mathrm{spad}}$ of 0.55--0.65 at 1550\,nm aligns with InGaAs SPAD modules coupled via telecom-grade fiber, and a timing jitter $\sigma_{\mathrm{spad}}$ of $\sim$50\,ps is attainable with picosecond-class time-tagging electronics combined with low-jitter SPAD detectors. These mappings ensure that the simulation framework reflects realistic design constraints and performance envelopes, bridging the gap between analytical modeling and field-deployable systems. In the remainder of this section, we examine the impact of key parameters (including spatial uncertainty \( \sigma_p \), beam waist \( w_z \), number of CCR elements \( N_{\mathrm{ar}} \), and background photon level \( \mu_{\mathrm{bg}} \)) on the accuracy of wireless quantum synchronization. The specific values used for each parameter sweep are indicated in the corresponding figure captions.

\begin{figure}
	\begin{center}
		\includegraphics[width=3.2 in, height =4.5 cm]{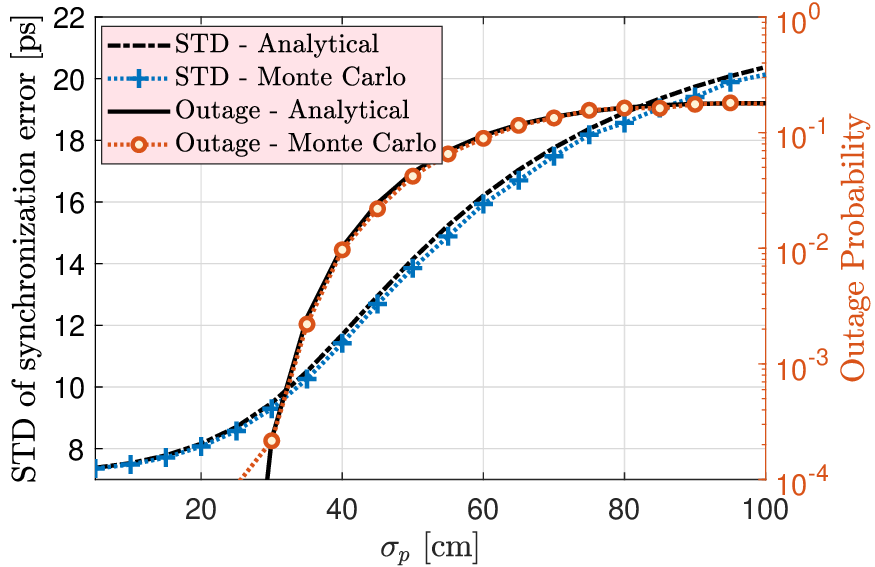}	
		\caption{Outage probability and synchronization timing error versus spatial uncertainty \( \sigma_p \).  Analytical results closely match Monte Carlo simulations.}
		\label{na1}
	\end{center}
\end{figure}

Fig.~\ref{cz} compares the analytical model of the photon count distribution \( \mathbb{P}(N_{\mathrm{sig}}) \) derived in Proposition~\ref{prop:N_sig} with Monte Carlo simulation results for different beam waists \( w_z \in \{25, 50, 75, 100\} \,\mathrm{cm} \). Despite the underlying system involving multiple dependent and independent random variables, the analytical expression, which requires only a two-dimensional integral, closely matches the simulation results, confirming its accuracy and computational efficiency.
The results also demonstrate the significant influence of \( w_z \) on the distribution. For smaller values such as \( w_z = 25\,\mathrm{cm} \), the distribution becomes wider with a higher probability of low photon counts due to sharper misalignment sensitivity. As \( w_z \) increases, the system becomes more robust to spatial deviations \( \mathbf{r}_{\mathrm{dev}} \), resulting in a narrower distribution but with a lower peak \( N_{\mathrm{sig}} \).

Fig.~\ref{na1} illustrates the impact of spatial uncertainty \( \sigma_p \) on two critical synchronization metrics: (i)~the outage probability of the wireless quantum synchronization process, and (ii)~the standard deviation (STD) of synchronization timing error. As shown, even small increases in \( \sigma_p \) sharply raise the outage probability, indicating a higher chance of \( N_{\mathrm{sig}} \) falling below the detection threshold. Note that, synchronization is declared unsuccessful if no valid detection occurs within the total duration of \( 100\,\mu\mathrm{s} \); the outage probability is defined as the likelihood of this event. In addition, the figure illustrates that the STD of synchronization timing error increases with \( \sigma_p \), indicating degraded temporal precision. These trends highlight the sensitivity of the system to spatial deviations. Notably, the analytical results are once again in excellent agreement with the Monte Carlo simulations, demonstrating that the proposed analytical expressions can accurately capture system behavior with significantly lower computational cost.

\begin{figure}
	\begin{center}
		\includegraphics[width=3 in, height =4.2 cm]{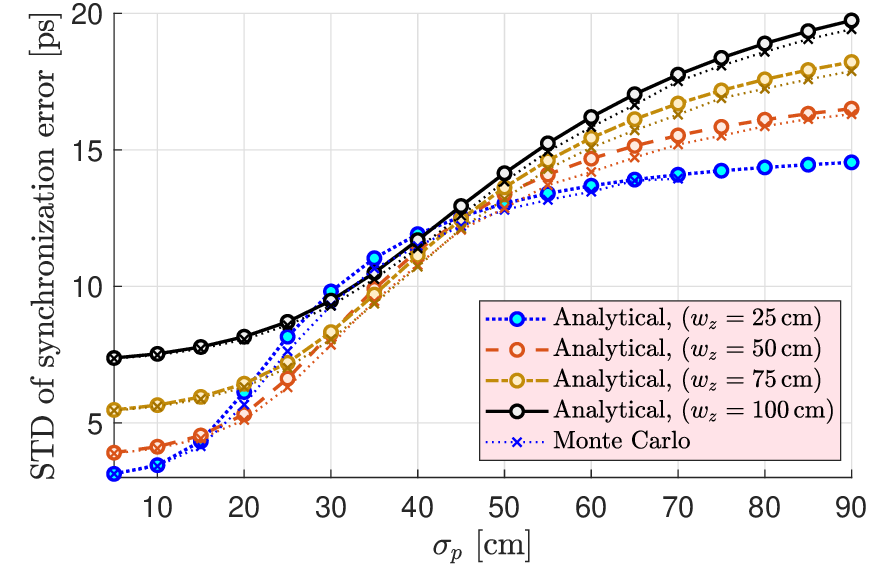}	
		\caption{Synchronization error versus \( \sigma_p \) for different beam waist values \( w_z \). The optimal \( w_z \) varies with \( \sigma_p \), emphasizing the need for adaptive selection based on current channel conditions.}	
		\label{na2}
	\end{center}
\end{figure}

Fig.~\ref{na2} explores the role of beam waist \( w_z \) as a tunable parameter in the synchronization process, evaluating its effect on synchronization error across different values of spatial uncertainty \( \sigma_p \). The results indicate that the optimal choice of \( w_z \) depends strongly on the level of misalignment, with no single setting being optimal under all conditions. Moreover, the optimal \( w_z \) is also influenced by other channel parameters, highlighting the need for adaptive configuration. Given its low computational complexity, the analytical model can be effectively employed to track channel variations and dynamically determine the optimal beam waist in real time.

\begin{figure}
	\begin{center}
		\includegraphics[width=3.0 in, height =4.2 cm]{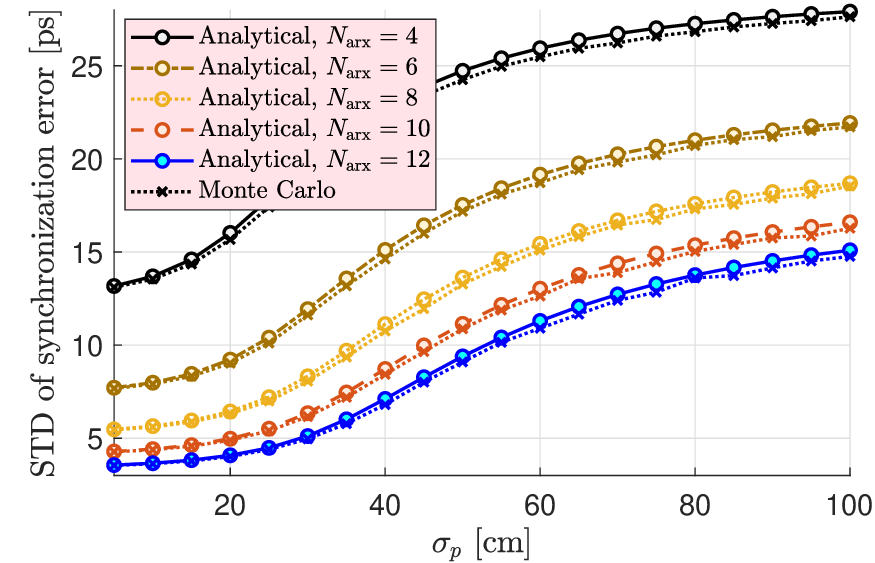}	
		\caption{Synchronization error versus \( \sigma_p \) for different CCR array sizes \( N_{\mathrm{ar}x} = N_{\mathrm{ar}y} \). Larger arrays improve accuracy by increasing photon counts.}
		\label{na3}
	\end{center}
\end{figure}

Fig.~\ref{na3} illustrates the STD of synchronization timing error as a function of \( \sigma_p \), for various square CCR array sizes with \( N_{\mathrm{ar}x} = N_{\mathrm{ar}y} \). As expected, increasing the number of CCR elements improves synchronization accuracy due to higher photon detection rates. However, the gain in accuracy exhibits diminishing returns: the improvement from \( N_{\mathrm{ar}x} = 4 \) to \( 8 \) is substantial, while the gain from \( 8 \) to \( 12 \) is more moderate. Although increasing \( N_{\mathrm{ar}} \) involves minimal hardware complexity due to the passive nature of CCRs, larger array sizes may impose physical constraints, especially in size-limited applications.

\begin{figure}
	\begin{center}
		\includegraphics[width=3.0 in, height =4.2 cm]{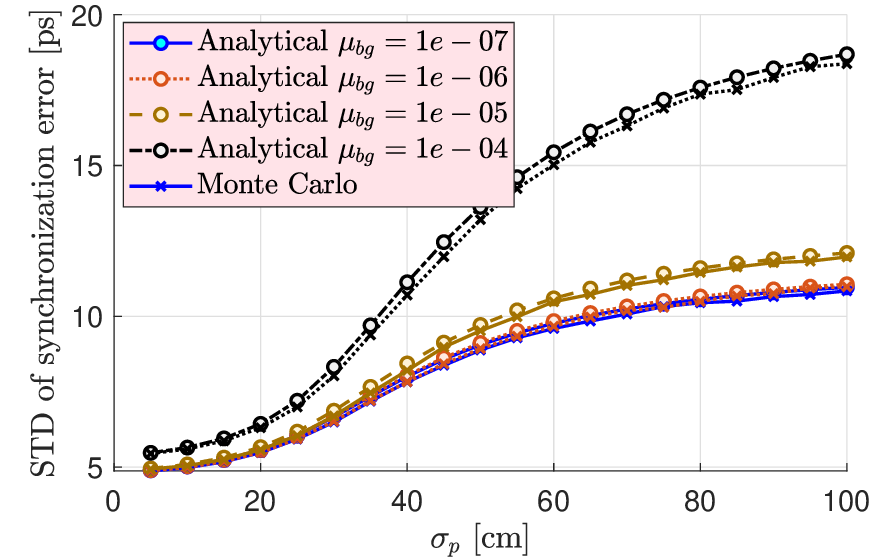}	
		\caption{Synchronization error versus background photon rate \( \mu_{\mathrm{bg}} \). Lower background noise improves accuracy up to a point, beyond which \( N_{\mathrm{bg}} \ll N_{\mathrm{sig}} \) and further gains saturate.}
		\label{na4}
	\end{center}
\end{figure}

Fig.~\ref{na4} illustrates the synchronization error as a function of \( \sigma_p \), for several values of background photon rate \( \mu_{\mathrm{bg}} \). The parameter \( \mu_{\mathrm{bg}} \) varies throughout the day depending on the sun’s position, and can be partially controlled through wavelength filtering or narrowing the field-of-view (FoV), albeit at the cost of increased system sensitivity to other imperfections such as angle-of-arrival errors.
At low \( \sigma_p \), the received signal photon count \( N_{\mathrm{sig}} \) is significantly higher than the background count \( N_{\mathrm{bg}} \), resulting in nearly identical performance across different \( \mu_{\mathrm{bg}} \) values. However, as \( \sigma_p \) increases, the probability of low \( N_{\mathrm{sig}} \) rises sharply, leading to increased synchronization error. This degradation is more pronounced at higher background levels, since the synchronization error component due to background fluctuations scales inversely with \( N_{\mathrm{sig}}^2 \) as observed in \eqref{eq:var_nch_final2}, while the timing jitter from SPADs scales with \( 1/N_{\mathrm{sig}} \). Therefore, for larger \( \mu_{\mathrm{bg}} \), synchronization performance deteriorates more rapidly as spatial uncertainty increases.

The simulations were performed for a 500\,m link, ensuring that the reported $>99$\% reliability and sub-5\,ps synchronization accuracy are attainable for shorter links as well. For longer links, achieving similar results would require more advanced hardware, such as sub-microradian beam-steering precision, which is currently challenging to obtain commercially. From a practical standpoint, deploying the proposed synchronization links also demands addressing key engineering factors: maintaining optical alignment tolerances on the order of sub-millimeters to preserve link availability, ensuring thermal stability of both transmitter and receiver optics to prevent slow drifts, and leveraging existing urban infrastructure to reduce deployment costs without compromising performance.

\textcolor{blue}{\section{Practical Implementation and Scalability Analysis}
\label{sec:practical}
This section addresses practical aspects of realizing the proposed synchronization architecture and discusses its scalability under real-world constraints.}

\textcolor{blue}{\subsection{Hardware Realization and Power Considerations}
\label{subsec:hardware}
The quantum hub requires an entangled-photon source, typically based on SPDC in a nonlinear crystal (e.g., PPKTP) pumped by a stabilized laser \cite{anwar2021entangled}. Single-photon detection is performed by InGaAs SPADs or SNSPDs \cite{fleming2025high}, coupled with high-speed time-tagging electronics. The optical output power is low (sub-mW), but the total electrical power consumption of the pump-laser driver, cooling, and readout circuits can reach several tens of watts. The target side remains completely passive, employing only a CCR array.}

\textcolor{blue}{State-of-the-art laboratory SPDC sources have demonstrated very high brightness and coincidence rates in the multi-MHz to tens-of-MHz regime, while internal pair-generation rates can be substantially higher \cite{brambila2023ultrabright}.
The GHz rate is therefore a forward-looking benchmark rather than a demonstrated operating point \cite{pang2025versatile}.
Recent advances in integrated quantum photonics, particularly resonantly enhanced microring-based nonlinear platforms,
have shown that high internal photon-pair generation rates and gigahertz-class repetition rates are technologically feasible,
suggesting a plausible path toward future GHz-rate entanglement distribution \cite{pang2025versatile,rahmouni2024entangled}.
Furthermore, recent demonstrations of high-dimensional time-bin entangled photons confirm the rapid maturation of robust, high-capacity quantum sources suitable for such deployments \cite{yu2025quantum}.}

\textcolor{blue}{\subsection{Multi-Pair Emission at the Quantum Hub}
\label{subsec:multipair_tx}
The entangled-photon source at the quantum hub is based on SPDC, which is inherently probabilistic and may occasionally generate multiple photon pairs within a single time slot. In the proposed framework, the source is operated in a low-mean-pair regime, such that the probability of multi-pair emission remains small. Although multi-pair events cannot be deterministically identified at the reference detector due to non-unit detection efficiency and residual losses, their impact is mitigated statistically by filtering ambiguous time bins during post-processing. As a result, multi-pair emissions contribute primarily to a small increase in accidental coincidences rather than causing systematic synchronization errors. If higher aggregate photon rates are required, standard multiplexing techniques across multiple low-brightness sources can be employed \cite{ma2011experimental}, reducing per-channel multi-pair probability while preserving overall system scalability.}

\textcolor{blue}{\subsection{Signal--Background Overlap and Background Mitigation}
\label{subsec:background_overlap}
Background photons constitute a major practical impairment in free-space single-photon links, particularly under daylight conditions.
In the proposed architecture, background contamination is mitigated through a combination of temporal, spectral, and spatial discrimination mechanisms. First, signal photons are generated within well-defined time bins, whereas background photons arrive randomly over the entire slot duration.
This temporal structure enables effective background suppression via narrow coincidence windows and time-gated detection \cite{gobby2004quantum}.
	Second, narrow-band spectral filtering around the operating wavelength (e.g., 1550~nm) significantly reduces out-of-band background radiation.}
	
\textcolor{blue}{Most importantly, the use of a CCR at the remote node enforces a deterministic round-trip propagation path for the signal photon \cite{10944637}.
	Due to the mutually orthogonal mirror structure of the CCR, any photon that is intercepted is reflected back along its incident direction \cite{10897286}.
	As a result, the angle-of-arrival uncertainty at the receiver is tightly bounded, primarily limited by diffraction and atmospheric effects.
	This property enables aggressive spatial filtering at the receiver, for example by coupling the return signal into a single-mode fiber with a microradian-scale field of view.
	Such strong spatial selectivity substantially suppresses background photons that do not follow the CCR-defined path.
}

\textcolor{blue}{\subsection{Robustness to Pointing Errors and CCR Imperfections}
\label{subsec:robustness}
CCRs inherently compensate for small angular misalignments and platform vibrations, because the three mutually perpendicular mirrors return the incident light along its original direction. This makes the link tolerant to typical mechanical jitter. The main practical degradation factor is surface contamination (dust, moisture), which reduces reflectivity. Protective coatings and periodic maintenance can mitigate this effect. Atmospheric turbulence, already incorporated via the Gamma-Gamma model, becomes more pronounced at longer ranges.}

\textcolor{blue}{\subsection{Background Suppression via Spatial Filtering}
\label{subsec:background}
The deterministic return path of the CCR-reflected photon enables aggressive spatial filtering. By coupling the receive aperture to a single-mode fiber (SMF), the effective field of view can be narrowed to match the diffraction-limited return beam. Combined with narrow-band spectral filtering (e.g., at 1550 nm) and temporal gating, this multi-stage filtering dramatically suppresses ambient background photons \cite{pirandola2020advances}. Consequently, \(\mu_{\mathrm{bg}}\) can be kept low through careful optical design.}

\textcolor{blue}{\subsection{Operational Trade-offs and Scalability Limits}
\label{subsec:scalability}
The single-photon nature of the link imposes a trade-off between channel gain and robustness to misalignment.
In our model, misalignment is captured by the spatial uncertainty parameter $\sigma_p$ in the target plane,
which is on the order of centimeters in the simulated scenarios.
In practice, $\sigma_p$ mainly originates from angular pointing and tracking errors at the transmitter and scales approximately as
$\sigma_p \approx L_{\mathrm{tar}}\sigma_\theta$, such that for a fixed pointing precision,
the effective misalignment increases with link distance.
Our results indicate that for centimeter-level spatial uncertainty, moderate turbulence
($C_n^2 \sim 10^{-15}\ \mathrm{m^{-2/3}}$) allows sub-nanosecond synchronization over links on the order of a few hundred meters.}

\textcolor{blue}{Scaling toward city-scale coverage therefore primarily requires improved beam-tracking accuracy
or reduced beam divergence at the central hub to limit the growth of spatial uncertainty with distance,
while keeping the remote target nodes fully passive and low-cost.
For smaller coverage radii, relaxed tracking requirements and correspondingly lower photon generation rates are sufficient,
whereas extending the operational range mainly necessitates tighter beam tracking and more careful system configuration.
At larger city scales, practical deployment will further depend on appropriate hub placement,
three-dimensional urban geometry, and line-of-sight availability,
suggesting that multi-hub layouts and environment-aware planning are key enablers for scalable operation.}

\appendices
\section{}\label{AppA}
Let \( w_z \) represent the beam waist of the Gaussian laser at the depth of the CCR array, which determines the spatial spread of the photon probability distribution in the \( x\text{--}y \) plane. The spatial PDF of detecting a photon at position \( \mathbf{r} = (x, y) \), when the beam is centered at \( \mathbf{p}(j_x, j_y) \), is given by the following two-dimensional circularly symmetric Gaussian distribution:
\begin{align}
	f_{\mathrm{ph}}(x, y; j) = \frac{2}{\pi w_z^2} \exp\left( -\frac{2 \| \mathbf{r} - \mathbf{p}(j_x, j_y) \|^2}{w_z^2} \right),
	\label{eq:gaussian_beam_pdf}
\end{align}
where \( \| \mathbf{r} - \mathbf{p}(j_x, j_y) \| \) denotes the Euclidean distance between the observation point and the beam center.

To evaluate the probability that an incident photon is captured and reflected by the $i$th CCR unit, located at position $\mathbf{p}_{\mathrm{ar},i} = (x_{\mathrm{ar},i}, y_{\mathrm{ar},i})$ in the $x$--$y$ plane, we integrate the spatial PDF of the beam over the effective aperture area $A_{\mathrm{ar},i}$ of the CCR as
\begin{align}
	P_{\mathrm{hit},i,j} = \frac{2}{\pi w_z^2}    \iint_{A_{\mathrm{ar},i}} 
	\exp\left( -\frac{2 \| \mathbf{r} - \mathbf{p}(j_x, j_y) \|^2}{w_z^2} \right)
	 dx\, dy,
	\label{eq:phit_integral}
\end{align}
The integration domain $A_{\mathrm{ar},i}$ represents the effective surface area of the $i$th CCR unit.

The integral in \eqref{eq:phit_integral} can be significantly simplified under practical considerations. In most realistic designs, instead of using a single large CCR—which suffers from increased depth, higher volume, mechanical constraints, and limited user-friendliness due to excessive size and weight, it is preferable to deploy an array of small CCR units. These arrays typically feature sub-elements on the order of $1~\mathrm{cm}^2$.
This modular structure not only reduces material and fabrication cost, but also allows flexible geometric configurations and better integration with compact platforms. Given that the beam waist $w_z$ is typically much larger than the individual CCR size (i.e., $A_{\mathrm{ar},i} \ll \pi w_z^2$), the variation of the Gaussian pattern over the surface of each CCR unit becomes negligible. Therefore, \eqref{eq:phit_integral} can be simplified as:
\begin{align}
	P_{\mathrm{hit},i,j} \approx \frac{2 A_{\mathrm{ar},i}}{\pi w_z^2}    
    \exp\left( -\frac{2 \| \mathbf{p}_{\mathrm{ar},i} - \mathbf{p}(j_x, j_y) \|^2}{w_z^2} \right).
	\label{eq:phit_approx}
\end{align} 

Although CCR units ideally reflect photons directly back toward their origin, their finite aperture truncates the incident electric field, inducing mild diffraction \cite{li2008calculation}. This effect modifies the divergence of the returned beam, which can be approximated as:
\begin{align}
	\theta_{\mathrm{dev},i} \approx {\lambda}/{\sqrt{A_{\mathrm{ar},i}}},
\end{align}
where \( \lambda \) is the photon wavelength, and \( A_{\mathrm{ar},i} \) is the effective area of the $i$th CCR unit.
Consequently, the returned beam acquires a new beam waist at the receiver plane, given approximately by \cite{li2008calculation}:
\begin{align}
	w_{z2,i} \approx L_{\mathrm{tar}} \cdot \theta_{\mathrm{dev},i} = {\lambda L_{\mathrm{tar}}}  / {\sqrt{A_{\mathrm{ar},i}}},
	\label{eq:wz2i}
\end{align}
Based on \eqref{eq:wz2i} the probability that the photon is captured within the receiver aperture of radius \( r_{\mathrm{ap}} \) is given by:
\begin{align}
	P_{\mathrm{ap},i,j} &= \iint_{r \leq r_{\mathrm{ap}}} \frac{2}{\pi} \cdot \frac{A_{\mathrm{ar},i}}{\lambda^2 L_{\mathrm{tar}}^2} \exp\left(-\frac{2 A_{\mathrm{ar},i}}{\lambda^2 L_{\mathrm{tar}}^2} r^2 \right) dx\,dy \nonumber \\
	&= 1 - \exp\left( -\frac{2 A_{\mathrm{ar},i}}{\lambda^2 L_{\mathrm{tar}}^2} \cdot r_{\mathrm{ap}}^2 \right).
	\label{eq:Pap_Aari}
\end{align}
Under the assumption that all CCR units have identical effective aperture area \( A_{\mathrm{ar},i} = A_{\mathrm{ar}} \), the capture probability of a reflected photon at the receiver becomes independent of both \( i \) and \( j \). Consequently, we define a uniform aperture capture probability:
\begin{align}
	P_{\mathrm{ap},i,j} = P_{\mathrm{ap}}, \quad \forall i,j.  \label{eqg1}
\end{align}
Finally by substituting \eqref{eq:phit_approx} and \eqref{eqg1} in \eqref{eq:photon_reception_i}, the PDF of $P_{\mathrm{hap},i,j}$ is derived in \eqref{eqg2}.

\section{}\label{AppB}
In this appendix, we derive the conditional probability distribution of the total single-photon reception probability $P_{\mathrm{rec},j}$ given the spatial offset vector $\mathbf{r}_{\mathrm{dev}}$.

The spatial coherence length $r_0$ of atmospheric turbulence in FSO channels depends on the optical wavelength $\lambda$ and the propagation path length $L_{\mathrm{tar}}$, and is given by \cite{andrews2005laser}:
\begin{align}
	r_0 = \left[ 0.423 \, k^2 \, C_n^2 \, L_{\mathrm{tar}} \right]^{-3/5},
	\label{eq:coherence_length}
\end{align}
where $k = 2\pi / \lambda$ is the optical wave number, and $C_n^2$ is the refractive index structure parameter (typically ranging from $10^{-17}$ to $10^{-13}~\mathrm{m}^{-2/3}$). For near-infrared FSO links with $\lambda \sim 1550$~nm and path lengths in the order of hundreds of meters, $r_0$ typically evaluates in order of cm.
In our system, the CCR array is arranged with inter-element spacing larger than the coherence length $r_0$, promoting statistical independence among the fading coefficients $\{ h_{\mathrm{a},i,j} \}$ and enabling spatial diversity gain, which in turn enhances the overall single-photon detection probability. 
Given the uncorrelated nature of the turbulence-induced fading coefficients $\{ h_{\mathrm{a},i,j} \}$, the total reception probability $P_{\mathrm{rec},j}$ in \eqref{eq:Precj_updated} can be expressed as a weighted sum of independent random variables. Specifically, we define the spatial weights:
\begin{align}
	w_{i,j} \triangleq \exp\left( -\frac{2 \| \mathbf{p}_{\mathrm{ar},i} - \mathbf{p}'(j_x, j_y) - \mathbf{r}_{\mathrm{dev}} \|^2}{w_z^2} \right),
\end{align}
and rewrite the reception probability as:
\begin{align}
	P_{\mathrm{rec},j} = c_0 \cdot \sum_{i=1}^{N_{\mathrm{ar}}} h_{\mathrm{a},i,j} \cdot w_{i,j},
\end{align}
where $c_0$ collects all deterministic scalar terms in \eqref{eq:Precj_updated}.

By the Central Limit Theorem, as $N_{\mathrm{ar}}$ grows large and the terms remain independent and non-dominant, the distribution of $P_{\mathrm{rec},j}$ conditioned on the spatial offset $\mathbf{r}_{\mathrm{dev}}$ converges to a Gaussian distribution:
\begin{align}
	P_{\mathrm{rec},j | \mathbf{r}_{\mathrm{dev}}}   \sim \mathcal{N}(\mu_j, \sigma_j^2), \label{eqr1}
\end{align}
with conditional mean and variance: 
\begin{align}
	\resizebox{0.99\hsize}{!}{$\mu_j = c_0 \cdot \sum_{i=1}^{N_{\mathrm{ar}}} \mathbb{E}[h_{\mathrm{a},i,j}] \cdot w_{i,j},~\&~~
	\sigma_j^2 = c_0^2 \cdot \sum_{i=1}^{N_{\mathrm{ar}}} \mathrm{Var}[h_{\mathrm{a},i,j}] \cdot w_{i,j}^2.$} \nonumber
\end{align}
For a Gamma-Gamma distributed fading coefficient with parameters $(\alpha, \beta)$, the mean and variance are given by \cite{andrews2005laser}:
\begin{align}
	\mathbb{E}[h_{\mathrm{a},i,j}] &= 1,~~ \&~~
	\mathrm{Var}[h_{\mathrm{a},i,j}] &= \left( \frac{1}{\alpha} + \frac{1}{\beta} + \frac{1}{\alpha\beta} \right).
\end{align}
Substituting these into the expressions for $\mu_j$ and $\sigma_j^2$ yields:
\begin{align} \label{eq:sigma_j_final}
	\resizebox{0.89\hsize}{!}{$ \mu_j = c_0 \cdot \sum_{i=1}^{N_{\mathrm{ar}}} w_{i,j}, ~~\&~~
	\sigma_j^2 = c_0^2 \cdot \left( \frac{1}{\alpha} + \frac{1}{\beta} + \frac{1}{\alpha\beta} \right)
	\cdot \sum_{i=1}^{N_{\mathrm{ar}}} w_{i,j}^2. $}
\end{align}
Finally, using \eqref{eqr1} and \eqref{eq:sigma_j_final}, the PDF of $P_{\mathrm{rec},j}$ conditioned on $\mathbf{r}_{\mathrm{dev}}$ is obtained in \eqref{eq:precj_pdf}.

\section{}\label{AppC}
In this appendix, we derive the variance of the quantum synchronization error $n_{\mathrm{ch}}$ defined in \eqref{eq:nch_model}, which captures the estimation noise in the recovered round-trip propagation delay $t_{\mathrm{ch}}$.
Let us define the random variables $X_{j'}$ for each valid detection slot $j' \in \{1, \dots, N_{\mathrm{tot}}\}$ as:
\begin{align}
	X_{j'} = 
	\begin{cases}
		t_{\mathrm{e,rx},j'} - t_{\mathrm{e,ref},j'}, & \text{if } s_{j'} = 0, \\
		t_{\mathrm{bg},j'} - t_{\mathrm{e,ref},j'},     & \text{if } s_{j'} = 1.
	\end{cases}
\end{align}
Hence, we can rewrite \eqref{eq:nch_model} compactly as:
\begin{align}
	n_\mathrm{ch} = \frac{1}{N_{\mathrm{tot}}} \sum_{j'=1}^{N_{\mathrm{tot}}} X_{j'}.
	\label{eq_appd_nch_xj}
\end{align}

We now compute $\mathrm{Var}(n_\mathrm{ch}) = \mathbb{E}[n_\mathrm{ch}^2] - \left(\mathbb{E}[n_\mathrm{ch}]\right)^2$. Since $n_\mathrm{ch}$ is a zero-mean random variable, $\mathbb{E}[n_\mathrm{ch}] = 0$, and hence:
\begin{align}
	\mathrm{Var}(n_\mathrm{ch}|N_{\mathrm{tot}}) = \mathbb{E}[n_\mathrm{ch}^2|N_{\mathrm{tot}}] = \frac{1}{N_{\mathrm{tot}}^2} \sum_{j'=1}^{N_{\mathrm{tot}}} \mathbb{E}[X_{j'}^2].
\end{align}

To compute $\mathbb{E}[X_{j'}^2]$, we consider the two cases:
\begin{itemize}
	\item For $s_{j'} = 0$, $X_{j'} = t_{\mathrm{e,rx},j'} - t_{\mathrm{e,ref},j'}$, which is a difference of two independent zero-mean Gaussian variables with variance $\sigma_{\mathrm{spad}}^2$, so:
	\[
	\mathbb{E}[X_{j'}^2 \mid s_{j'} = 0] = 2\sigma_{\mathrm{spad}}^2.
	\]
	
	\item For $s_{j'} = 1$, the error term is defined as
	\[
	X_{j'} = t_{\mathrm{bg},j'} - t_{\mathrm{e,ref},j'},
	\]
	where $t_{\mathrm{bg},j'} \sim \mathcal{U}[-t_{\mathrm{qb}}/2,~ t_{\mathrm{qb}}/2]$ is the arrival time of a background photon, assumed to be uniformly distributed around the central point of each qubit slot,
	and $t_{\mathrm{e,ref},j'} \sim \mathcal{N}(0, \sigma_{\mathrm{spad}}^2)$ is the reference SPAD jitter, which is independent. 
	The variance of a uniform distribution \cite{papoulis2002probability}:
	\begin{align}
		\text{If } X \sim \mathcal{U}[a,b], \quad \mathrm{Var}(X) = \frac{(b - a)^2}{12}.
		\label{eq:var_uniform}
	\end{align}
	In our case, $a = -t_{\mathrm{qb}}/2$, $b = t_{\mathrm{qb}}/2$, which leads to $\mathrm{Var}(t_{\mathrm{bg}}) = t_{\mathrm{qb}}^2 / 12$.
	Therefore, the conditional second moment becomes
	\begin{align} \label{eq:var_uniform_plus_gaussian}
	\resizebox{0.79\hsize}{!}{$ \mathbb{E}[X_{j'}^2 \mid s_{j'} = 1] = \mathrm{Var}(t_{\mathrm{bg}}) + \mathrm{Var}(t_{\mathrm{e,ref}}) 
	= \frac{(t_{\mathrm{qb}})^2}{12} + \sigma_{\mathrm{spad}}^2, $}
   \end{align}	
\end{itemize}

\textcolor{blue}{Let $p_{\mathrm{bg}} = \frac{N_{\mathrm{bg}}}{N_{\mathrm{tot}}}$ denote the (random) background photon ratio. Then, taking expectations over $s_{j'}$ conditioned on $N_{\mathrm{sig}}$ and $N_{\mathrm{bg}}$:
	\begin{align}
		\mathbb{E}[X_{j'}^2 \mid N_{\mathrm{sig}}, N_{\mathrm{bg}}] &= (1 - p_{\mathrm{bg}}) \cdot 2\sigma_{\mathrm{spad}}^2  \nonumber \\ &~~~
		+ p_{\mathrm{bg}} \cdot \left( {t_{\mathrm{qb}}^2}/{12} + \sigma_{\mathrm{spad}}^2 \right).
	\end{align}
	Summing over $j'$ and normalizing by $N_{\mathrm{tot}}^2$, we obtain the conditional variance of the synchronization error given $N_{\mathrm{sig}}$ and $N_{\mathrm{bg}}$:
	\begin{align}
		&\resizebox{0.99\hsize}{!}{$ \mathrm{Var}(n_\mathrm{ch} \mid N_{\mathrm{sig}}, N_{\mathrm{bg}})    
			=\frac{1}{N_{\mathrm{tot}}} \left[ 2\sigma_{\mathrm{spad}}^2 (1 - p_{\mathrm{bg}}) 
			+ \left( \frac{t_{\mathrm{qb}}^2}{12} + \sigma_{\mathrm{spad}}^2 \right) p_{\mathrm{bg}} \right] $}\nonumber \\
		&= \frac{1}{N_{\mathrm{tot}}} \left[ \sigma_{\mathrm{spad}}^2 (1 + p_{\mathrm{bg}}) + {t_{\mathrm{qb}}^2}/{12} \cdot p_{\mathrm{bg}} \right].
		\label{eq:var_nch_final}
\end{align} }

\textcolor{blue}{Note that both $N_{\mathrm{sig}}$ and $N_{\mathrm{bg}}$ are random variables. Therefore, the conditional variance $\mathrm{Var}(n_{\mathrm{ch}} \mid N_{\mathrm{sig}}, N_{\mathrm{bg}})$ is itself random. Since synchronization is attempted only when $N_{\mathrm{sig}} \ge N_{\mathrm{t,min}}$, we evaluate the expected conditional variance over all realizations that satisfy this detection threshold. This expected conditional variance is given by:
	\begin{align}
		&\overline{\mathrm{Var}}(n_{\mathrm{ch}}) 
		= \sum_{N_{\mathrm{sig}}=N_{\mathrm{t,min}}}^{N_{\mathrm{s,max}}} 
		\sum_{N_{\mathrm{bg}}=\max(0, N_{\mathrm{t,min}}-N_{\mathrm{sig}})}^{N_{\mathrm{b,max}}} 
		\mathbb{P}(N_{\mathrm{sig}}) \nonumber \\ &~~~\times \mathbb{P}(N_{\mathrm{bg}})  
		\mathrm{Var}(n_{\mathrm{ch}} \mid N_{\mathrm{sig}}, N_{\mathrm{bg}}) ,
		\label{eq:expected_var_nch}
	\end{align}
	where $N_{\mathrm{t,min}}$ is the minimum required number of valid (non‑null) photon detections needed to perform meaningful synchronization, and $N_{\mathrm{s,max}}$ and $N_{\mathrm{b,max}}$ denote upper bounds on signal and background counts based on channel parameters. The joint probability mass function is assumed to factor as $\mathbb{P}(N_{\mathrm{sig}}, N_{\mathrm{bg}}) = \mathbb{P}(N_{\mathrm{sig}}) \cdot \mathbb{P}(N_{\mathrm{bg}})$ under the assumption of statistical independence between signal and background processes.
	Using \eqref{eq:var_nch_final} and \eqref{eq:expected_var_nch}, the expected conditional variance of the synchronization error is obtained in \eqref{eq:var_nch_final2}.}

\section{} \label{AppD}
To derive the expected number of signal photon detections \( \mathbb{E}[N_{\mathrm{sig}}] \), we first express the total count as a sum over all spatial grid cells:
$N_{\mathrm{sig}} = \sum_{j=1}^{N_{\mathrm{gr}}} N_{\mathrm{sig},j}$,
where \(N_{\mathrm{sig},j}\) denotes the number of detected signal photons corresponding to the \(j\)th grid cell.
Each grid cell contains \(L_{\mathrm{sv}}\) qubit intervals, as defined in \eqref{eq:L_sv}. Only intervals in which exactly one photon pair is generated lead to valid (non-null) entries in the reference and received vectors. 
Each time slot in grid cell \(j\) contributes to the signal photon count if (i) a single photon pair is generated, and (ii) the photon is successfully detected. Based on \eqref{eq:poisson} and \eqref{eqg6}, these two events occur independently. \textcolor{blue}{The probability of generating exactly one photon pair is \(\mu_t e^{-\mu_t}\).  
	For a given channel realization (i.e., fixed atmospheric turbulence and pointing misalignment), the channel gain \(\mu_{\mathrm{ch},j}\) is a deterministic coefficient. Under the low‑rate regime \(\mu_{\mathrm{ch},j} \ll 1\), the probability of detecting exactly one photon in that slot is approximately \(\mu_{\mathrm{ch},j}\). Thus, the total success probability per slot conditioned on \(\mu_{\mathrm{ch},j}\) is:}
\[
p_j = \mu_t e^{-\mu_t} \cdot \mu_{\mathrm{ch},j}.
\]
Assuming \(L_{\mathrm{sv}}\) independent trials per grid, the probability that exactly \(n\) photons are successfully detected in grid \(j\) conditioned on $\mu_{\mathrm{ch},j}$ is given by:
\begin{align}
	\mathbb{P}(N_{\mathrm{sig},j} = n \mid \mu_{\mathrm{ch},j}) = \binom{L_{\mathrm{sv}}}{n} p_j^n (1 - p_j)^{L_{\mathrm{sv}} - n}, \label{eq:bin_exp}
\end{align}
which holds for \(n = 0, 1, \dots, L_{\mathrm{sv}}\).

Given that the considered quantum channel is highly lossy (\(\mu_{\mathrm{ch},j} \ll 1\)) and the average photon pair generation rate is sub-unity (\(\mu_t < 1\)), the success probability per slot \(p_j = \mu_t e^{-\mu_t} \mu_{\mathrm{ch},j}\) is much smaller than one. Under these conditions, the binomial distribution in \eqref{eq:bin_exp} is well approximated by a Poisson distribution \cite[Sec.~6.5.2]{shynk2012probability}:
\begin{align}
	\resizebox{0.89\hsize}{!}{$ \mathbb{P}(N_{\mathrm{sig},j} = n|\mu_{\mathrm{ch},j}) \approx \frac{(\lambda_t \mu_{\mathrm{ch},j})^n}{n!} e^{-\lambda_t \mu_{\mathrm{ch},j}}, \quad \lambda_t = L_{\mathrm{sv}} \mu_t e^{-\mu_t} . $} \label{eq:poisson_approx}
\end{align}
Since each \(N_{\mathrm{sig},j}\) is conditionally independent and follows a Poisson distribution with mean \(\lambda_t \mu_{\mathrm{ch},j}\), the total signal count \(N_{\mathrm{sig}}\) conditioned on ${\mu_{\mathrm{ch},j}}$ also follows a Poisson distribution with aggregated mean \cite{shynk2012probability}:
\begin{align}
	\mathbb{P}(N_{\mathrm{sig}} = n \mid \{\mu_{\mathrm{ch},j}\}) = 
	\frac{(\lambda_t \mu'_{\mathrm{ch}})^n\exp\left(-\lambda_t \mu'_{\mathrm{ch}} \right)      }{n!} ,
	\label{eq:Nsig_poisson_total}
\end{align}
where 
$\mu'_{\mathrm{ch}} = 
	\sum_{j=1}^{N_{\mathrm{gr}}} \mu_{\mathrm{ch},j},
	\label{mu_ch_t}$.
Using \eqref{eq:precj_pdf} and \eqref{eqg7}, the conditional density of \(\mu_{\mathrm{ch},j}\) given \(\mathbf{r}_{\mathrm{dev}}\) is:
\begin{align}
	&f_{\mu_{\mathrm{ch},j} | \mathbf{r}_{\mathrm{dev}}}(\mu) = 
	\frac{1}{\eta_{\mathrm{spad}} \sqrt{2\pi c_0^2 c_{\alpha,\beta} \sum_{i=1}^{N_{\mathrm{ar}}} w_{i,j,\mathbf{r}_{\mathrm{dev}}}^2 }} \nonumber \\
	&\quad \times \exp\left( -\frac{\left( \frac{\mu}{\eta_{\mathrm{spad}}} - c_0 \sum_{i=1}^{N_{\mathrm{ar}}} w_{i,j,\mathbf{r}_{\mathrm{dev}}} \right)^2}
	{2c_0^2 c_{\alpha,\beta} \sum_{i=1}^{N_{\mathrm{ar}}} w_{i,j,\mathbf{r}_{\mathrm{dev}}}^2} \right).
	\label{eq:mu_ch_cond_pdf_explicit}
\end{align} 
Based on \eqref{eq:mu_ch_cond_pdf_explicit}, since each $\mu_{\mathrm{ch},j} $ conditioned on $ \mathbf{r}_{\mathrm{dev}}$ follows a Gaussian distribution, and $\mu_{\mathrm{ch},j} $s conditioned on $ \mathbf{r}_{\mathrm{dev}}$ are independent, the sum
$\mu'_{\mathrm{ch}} = \sum_{j=1}^{N_{\mathrm{gr}}} \mu_{\mathrm{ch},j}$
is also Gaussian distributed as:
\begin{align}
	&f_{\mu'_{\mathrm{ch}} | \mathbf{r}_{\mathrm{dev}}}(\mu) = 
	\frac{1}{\sqrt{2\pi \eta_{\mathrm{spad}}^2 c_0^2 c_{\alpha,\beta} \sum_{j=1}^{N_{\mathrm{gr}}} \sum_{i=1}^{N_{\mathrm{ar}}} w_{i,j,\mathbf{r}_{\mathrm{dev}}}^2 }} \nonumber \\
	&\quad \times \exp\left( -\frac{\left( \mu - \eta_{\mathrm{spad}} c_0 \sum_{j=1}^{N_{\mathrm{gr}}} \sum_{i=1}^{N_{\mathrm{ar}}} w_{i,j,\mathbf{r}_{\mathrm{dev}}} \right)^2}
	{2 \eta_{\mathrm{spad}}^2 c_0^2 c_{\alpha,\beta} \sum_{j=1}^{N_{\mathrm{gr}}} \sum_{i=1}^{N_{\mathrm{ar}}} w_{i,j,\mathbf{r}_{\mathrm{dev}}}^2} \right).
	\label{eq:mu_ch_prime_pdf_final}
\end{align}

To obtain the unconditional distribution of $\mu'_{\mathrm{ch}}$, we marginalize over the spatial offset vector $\mathbf{r}_{\mathrm{dev}} = (x_{\mathrm{dev}}, y_{\mathrm{dev}})$, which follows a bivariate Gaussian distribution with zero mean and isotropic variance $\sigma_p^2$. The unconditional PDF is thus expressed as:
\begin{align}
	\label{f_mu1}
	\resizebox{0.89\hsize}{!}{$ f_{\mu'_{\mathrm{ch}}}(\mu) = \iint f_{\mu'_{\mathrm{ch}} | \mathbf{r}_{\mathrm{dev}}}(\mu) 
	f_{\mathbf{r}_{\mathrm{dev}}}(x_{\mathrm{dev}}, y_{\mathrm{dev}})~dx_{\mathrm{dev}} dy_{\mathrm{dev}}. $}
\end{align}
However, due to the nonlinear dependence of $\mu'_{\mathrm{ch}}$ on both $x_{\mathrm{dev}}$ and $y_{\mathrm{dev}}$ through the spatial weights $w_{i,j,\mathbf{r}_{\mathrm{dev}}}$, direct evaluation of \eqref{f_mu1} is analytically intractable. To facilitate tractable analysis while maintaining conservative accuracy, we adopt a worst-case approximation that the spatial offset occurs along only one axis (e.g., $x$-axis), while the other component remains fixed. Specifically, we set:
$x_{\mathrm{dev}} = r, \quad y_{\mathrm{dev}} = 0$,
which corresponds to a radial deviation of magnitude $r$ purely along the $x$-direction. Under this assumption, the 2D Gaussian PDF of $\mathbf{r}_{\mathrm{dev}}$ reduces to a Rayleigh distribution for $r$ with scale parameter $\sigma_p$:
\begin{align}
	\label{f_r_dev}
	f_{\mathbf{r}_{\mathrm{dev}}}(r) = \frac{r}{\sigma_p^2} \exp\left( -\frac{r^2}{2\sigma_p^2} \right), \quad r \geq 0.
\end{align}
Accordingly, the integral in \eqref{f_mu} simplifies to:
\begin{align}
	\label{f_mu3}
	f_{\mu'_{\mathrm{ch}}}(\mu) = \int_0^\infty f_{\mu'_{\mathrm{ch}} | \mathbf{r}_{\mathrm{dev}} = (r,0)}(\mu) 
	f_{\mathbf{r}_{\mathrm{dev}}}(r)~dr,
\end{align}
which allows efficient numerical evaluation while still capturing the degradation effect of radial misalignment in a worst-case scenario.

Finally using \eqref{eq:Nsig_poisson_total}, \eqref{eq:mu_ch_prime_pdf_final}, and \eqref{f_mu3}, the pdf of $N_{\mathrm{sig}}$ conditioned on $\mathbf{r}_{\mathrm{dev}}$ is provided in \eqref{prop:N_sig}.

\bibliographystyle{IEEEtran}
\balance
\bibliography{IEEEabrv,myref}

\begin{thebibliography}{10}
\providecommand{\url}[1]{#1}
\csname url@samestyle\endcsname
\providecommand{\newblock}{\relax}
\providecommand{\bibinfo}[2]{#2}
\providecommand{\BIBentrySTDinterwordspacing}{\spaceskip=0pt\relax}
\providecommand{\BIBentryALTinterwordstretchfactor}{4}
\providecommand{\BIBentryALTinterwordspacing}{\spaceskip=\fontdimen2\font plus
\BIBentryALTinterwordstretchfactor\fontdimen3\font minus
  \fontdimen4\font\relax}
\providecommand{\BIBforeignlanguage}[2]{{%
\expandafter\ifx\csname l@#1\endcsname\relax
\typeout{** WARNING: IEEEtran.bst: No hyphenation pattern has been}%
\typeout{** loaded for the language `#1'. Using the pattern for}%
\typeout{** the default language instead.}%
\else
\language=\csname l@#1\endcsname
\fi
#2}}
\providecommand{\BIBdecl}{\relax}
\BIBdecl

\bibitem{10487780}
S.~F. Bush, C.~G. Iversen, and W.~A. Challener, ``Design for high-precision
  time-sensitive networking: Synchronization for the quantum network control
  plane,'' \emph{IEEE Journal on Selected Areas in Communications}, vol.~42,
  no.~7, pp. 1861--1870, 2024.

\bibitem{colombo2022time}
S.~Colombo, E.~Pedrozo-Penafiel, A.~F. Adiyatullin, Z.~Li, E.~Mendez, C.~Shu,
  and V.~Vuleti{\'c}, ``Time-reversal-based quantum metrology with many-body
  entangled states,'' \emph{Nature Physics}, vol.~18, no.~8, pp. 925--930,
  2022.

\bibitem{zhu2022gnss}
L.~Zhu, H.~Zhang, X.~Li, F.~Zhu, and Y.~Liu, ``{GNSS timing performance
  assessment and results analysis},'' \emph{Sensors}, vol.~22, no.~7, p. 2486,
  2022.

\bibitem{tavella2020precise}
P.~Tavella and G.~Petit, ``Precise time scales and navigation systems: Mutual
  benefits of timekeeping and positioning,'' \emph{Satellite Navigation},
  vol.~1, no.~1, p.~10, 2020.

\bibitem{minetto2024nanosecond}
A.~Minetto, B.~Rat, M.~Pini, B.~D. Polidori, I.~De~Francesca, L.~C. Murillo,
  and F.~Dovis, ``{Nanosecond-level resilient GNSS-based time synchronization
  in telecommunication networks through WR-PTP HA},'' \emph{IEEE Systems
  Journal}, vol.~18, no.~1, pp. 327--338, 2024.

\bibitem{patoli2025fpga}
A.~A. Patoli and G.~Fortino, ``{FPGA-Based System Implementation of IEEE 1588
  Precision Time Protocol: A Review},'' \emph{IEEE Sensors Journal}, 2025.

\bibitem{gutierrez2021enhancing}
J.~L. Guti{\'e}rrez-Rivas, F.~Torres-Gonz{\'a}lez, E.~Ros, and J.~D{\'\i}az,
  ``{Enhancing white rabbit synchronization stability and scalability using P2P
  transparent and hybrid clocks},'' \emph{IEEE Transactions on Industrial
  Informatics}, vol.~17, no.~11, pp. 7316--7324, 2021.

\bibitem{sliwczynski2020picoseconds}
{\L}.~{\'S}liwczy{\'n}ski, P.~Krehlik, {\L}.~Buczek, and H.~Schnatz,
  ``Picoseconds-accurate fiber-optic time transfer with relative stabilization
  of lasers wavelengths,'' \emph{Journal of Lightwave Technology}, vol.~38,
  no.~18, pp. 5056--5063, 2020.

\bibitem{nande2025integrating}
S.~S. Nande, M.~I. Habibie, M.~Ghadimi, A.~Garbugli, K.~Kondepu, R.~Bassoli,
  and F.~H. Fitzek, ``Integrating quantum synchronization in future generation
  networks,'' \emph{Scientific Reports}, vol.~15, no.~1, p. 7617, 2025.

\bibitem{barhoumi2025quantum}
M.~Barhoumi, ``Quantum-assisted network time synchronisation: A literature
  review, considering examples and challenges,'' \emph{Available at SSRN
  5170022}, 2025.

\bibitem{spiess2023clock}
C.~Spiess, S.~T{\"o}pfer, S.~Sharma, A.~Kr{\v{z}}i{\v{c}}, M.~Cabrejo-Ponce,
  U.~Chandrashekara, N.~L. D{\"o}ll, D.~Riel{\"a}nder, and F.~Steinlechner,
  ``Clock synchronization with correlated photons,'' \emph{Physical Review
  Applied}, vol.~19, no.~5, p. 054082, 2023.

\bibitem{chin2022digital}
H.-M. Chin, N.~Jain, U.~L. Andersen, D.~Zibar, and T.~Gehring, ``Digital
  synchronization for continuous-variable quantum key distribution,''
  \emph{Quantum Science and Technology}, vol.~7, no.~4, p. 045006, 2022.

\bibitem{morales2019survey}
R.~Morales-Ferre, P.~Richter, E.~Falletti, A.~De~La~Fuente, and E.~S. Lohan,
  ``A survey on coping with intentional interference in satellite navigation
  for manned and unmanned aircraft,'' \emph{IEEE Communications Surveys \&
  Tutorials}, vol.~22, no.~1, pp. 249--291, 2019.

\bibitem{dabiri2022pointing}
M.~T. Dabiri and M.~Hasna, ``{Pointing error modeling of mmWave to THz
  high-directional antenna arrays},'' \emph{IEEE Wireless Communications
  Letters}, vol.~11, no.~11, pp. 2435--2439, 2022.

\bibitem{dabiri2018channel}
M.~T. Dabiri, S.~M.~S. Sadough, and M.~A. Khalighi, ``{Channel modeling and
  parameter optimization for hovering UAV-based free-space optical links},''
  \emph{IEEE Journal on Selected Areas in Communications}, vol.~36, no.~9, pp.
  2104--2113, 2018.

\bibitem{savino2024robust}
N.~Savino, J.~Leamer, R.~Saripalli, W.~Zhang, D.~Bondar, and R.~Glasser,
  ``Robust free-space optical communication utilizing polarization for the
  advancement of quantum communication,'' \emph{Entropy}, vol.~26, no.~4, p.
  309, 2024.

\bibitem{gobby2004quantum}
C.~Gobby, a.~Yuan, and A.~Shields, ``Quantum key distribution over 122 km of
  standard telecom fiber,'' \emph{Applied Physics Letters}, vol.~84, no.~19,
  pp. 3762--3764, 2004.

\bibitem{liu2019solid}
J.~Liu, R.~Su, Y.~Wei, B.~Yao, S.~F. C.~d. Silva, Y.~Yu, J.~Iles-Smith,
  K.~Srinivasan, A.~Rastelli, J.~Li \emph{et~al.}, ``A solid-state source of
  strongly entangled photon pairs with high brightness and
  indistinguishability,'' \emph{Nature nanotechnology}, vol.~14, no.~6, pp.
  586--593, 2019.

\bibitem{andrews2005laser}
L.~C. Andrews and R.~L. Phillips, ``Laser beam propagation through random
  media,'' \emph{Laser Beam Propagation Through Random Media: Second Edition},
  2005.

\bibitem{dabiri2025unified}
M.~T. Dabiri, M.~Hasna, S.~Al-Kuwari, and K.~Qaraqe, ``{A Unified Framework for
  UAV-Based Free-Space Quantum Links: Beam Shaping and Adaptive Field-of-View
  Control},'' \emph{arXiv e-prints}, pp. arXiv--2506, 2025.

\bibitem{shynk2012probability}
J.~J. Shynk, \emph{Probability, random variables, and random processes: theory
  and signal processing applications}.\hskip 1em plus 0.5em minus 0.4em\relax
  John Wiley \& Sons, 2012.

\bibitem{lyu2023nanomotion}
W.~Lyu, W.~Tang, W.~Yan, and M.~Qiu, ``Nanomotion of micro-objects driven by
  light-induced elastic waves on solid interfaces,'' \emph{Physical Review
  Applied}, vol.~19, no.~2, p. 024049, 2023.

\bibitem{liao2017satellite}
S.-K. Liao, W.-Q. Cai, W.-Y. Liu, L.~Zhang, Y.~Li, J.-G. Ren, J.~Yin, Q.~Shen,
  Y.~Cao, Z.-P. Li \emph{et~al.}, ``Satellite-to-ground quantum key
  distribution,'' \emph{Nature}, vol. 549, no. 7670, pp. 43--47, 2017.

\bibitem{anwar2021entangled}
A.~Anwar, C.~Perumangatt, F.~Steinlechner, T.~Jennewein, and A.~Ling,
  ``Entangled photon-pair sources based on three-wave mixing in bulk
  crystals,'' \emph{Review of Scientific Instruments}, vol.~92, no.~4, 2021.

\bibitem{fleming2025high}
F.~Fleming, W.~McCutcheon, E.~E. Wollman, A.~D. Beyer, V.~Anant, B.~Korzh,
  J.~P. Allmaras, L.~Narv{\'a}ez, S.~Leedumrongwatthanakun, G.~S. Buller
  \emph{et~al.}, ``{High-efficiency, high-count-rate 2D superconducting
  nanowire single-photon detector array},'' \emph{Optics Express}, vol.~33,
  no.~13, pp. 27\,602--27\,614, 2025.

\bibitem{brambila2023ultrabright}
E.~Brambila, R.~G{\'o}mez, R.~Fazili, M.~Gr{\"a}fe, and F.~Steinlechner,
  ``Ultrabright polarization-entangled photon pair source for
  frequency-multiplexed quantum communication in free-space,'' \emph{Optics
  Express}, vol.~31, no.~10, pp. 16\,107--16\,117, 2023.

\bibitem{pang2025versatile}
Y.~Pang, J.~E. Castro, T.~J. Steiner, L.~Duan, N.~Tagliavacche, M.~Borghi,
  L.~Thiel, N.~Lewis, J.~E. Bowers, M.~Liscidini \emph{et~al.}, ``{A Versatile
  Chip-Scale Platform for High-Rate Entanglement Generation with an AlGaAs
  Microresonator Array},'' in \emph{2025 Conference on Lasers and
  Electro-Optics (CLEO)}.\hskip 1em plus 0.5em minus 0.4em\relax IEEE, 2025,
  pp. 1--2.

\bibitem{rahmouni2024entangled}
A.~Rahmouni, R.~Wang, J.~Li, X.~Tang, T.~Gerrits, O.~Slattery, Q.~Li, and
  L.~Ma, ``{Entangled photon pair generation in an integrated SiC platform},''
  \emph{Light: Science \& Applications}, vol.~13, no.~1, p. 110, 2024.

\bibitem{yu2025quantum}
H.~Yu, S.~Sciara, M.~Chemnitz, N.~Montaut, B.~Crockett, B.~Fischer, R.~Helsten,
  B.~Wetzel, T.~A. Goebel, R.~G. Kr{\"a}mer \emph{et~al.}, ``Quantum key
  distribution implemented with d-level time-bin entangled photons,''
  \emph{Nature Communications}, vol.~16, no.~1, p. 171, 2025.

\bibitem{ma2011experimental}
X.-s. Ma, S.~Zotter, J.~Kofler, T.~Jennewein, and A.~Zeilinger, ``Experimental
  generation of single photons via active multiplexing,'' \emph{Physical Review
  A—Atomic, Molecular, and Optical Physics}, vol.~83, no.~4, p. 043814, 2011.

\bibitem{10944637}
M.~T. Dabiri, M.~Hasna, S.~Althunibat, and K.~Qaraqe, ``{UAV-BASED Dynamic FSO
  Access Networks: Technological Comparison, Design Considerations, and Future
  Directions},'' \emph{IEEE Wireless Communications}, vol.~32, no.~2, pp.
  247--253, 2025.

\bibitem{10897286}
M.~T. Dabiri and M.~Hasna, ``{A Novel MRR-UAV-Based Relay With Optical Network
  Coding: A Comparative Study With Optical IRS and Conventional UAV
  Relaying},'' \emph{IEEE Journal on Selected Areas in Communications},
  vol.~43, no.~5, pp. 1607--1620, 2025.

\bibitem{pirandola2020advances}
S.~Pirandola, U.~L. Andersen, L.~Banchi, M.~Berta, D.~Bunandar, R.~Colbeck,
  D.~Englund, T.~Gehring, C.~Lupo, C.~Ottaviani \emph{et~al.}, ``Advances in
  quantum cryptography,'' \emph{Advances in optics and photonics}, vol.~12,
  no.~4, pp. 1012--1236, 2020.

\bibitem{li2008calculation}
S.~Li, B.~Tang, and H.~Zhou, ``Calculation on diffraction aperture of cube
  corner retroreflector,'' \emph{Chinese Optics Letters}, vol.~6, no.~11, pp.
  833--836, 2008.

\bibitem{papoulis2002probability}
A.~Papoulis and S.~U. Pillai, \emph{{Probability, Random Variables, and
  Stochastic Processes}}, 4th~ed.\hskip 1em plus 0.5em minus 0.4em\relax
  McGraw-Hill, 2002.

\end{thebibliography}

\end{document}